\documentclass[11pt]{article}

\PassOptionsToPackage{table}{xcolor}

\usepackage[final]{acl}
\usepackage{times}
\usepackage{latexsym}
\usepackage[T1]{fontenc}
\usepackage[utf8]{inputenc}
\usepackage{microtype}
\usepackage{courier}  % Courier mono pairs better with Times body than cmtt
\makeatletter
\let\origtexttt\texttt
\renewcommand{\texttt}[1]{{\@tempdima=\f@size pt\multiply\@tempdima by 91\divide\@tempdima by 100
  \fontsize{\@tempdima}{1.2\@tempdima}\selectfont\origtexttt{#1}}}
\makeatother

\usepackage{hyperref}
\usepackage{url}
\usepackage{xurl}
\usepackage{booktabs}
\usepackage{tabularx}
\usepackage{multirow}
\usepackage{xcolor}
\usepackage{amsfonts}
\usepackage{amsmath}
\usepackage{nicefrac}
\usepackage{xcolor}   % `table` option forwarded above via \PassOptionsToPackage
\usepackage{graphicx}
\usepackage{enumitem}
\usepackage{float}
\usepackage[normalem]{ulem} % \sout for visible revision deletions

\usepackage{listings}
\usepackage{pifont}      % \ding{51} (check) / \ding{55} (cross)
\usepackage{tcolorbox}   % \begin{tcolorbox} for "Answer to RQ" summary boxes

\definecolor{hintbg}{RGB}{255,243,205}     % soft yellow for the hint line
\definecolor{kw}{RGB}{0,90,156}            % keyword blue
\definecolor{cmt}{RGB}{120,120,120}        % comment grey
\definecolor{str}{RGB}{170,55,55}          % string red
\definecolor{badbg}{RGB}{253,232,232}      % light red panel bg
\definecolor{goodbg}{RGB}{230,247,232}     % light green panel bg
\definecolor{good}{RGB}{30,120,40}         % darker green for marks
\definecolor{bad}{RGB}{170,40,40}          % darker red for marks

\lstdefinestyle{cppcase}{%
  basicstyle=\ttfamily\scriptsize,
  language=C++,
  morekeywords={vector,string,size_t,uint8_t,int32_t,int64_t,uint32_t,uint64_t,
                undefined4,undefined8,ulong,ushort,uchar,longlong,FUN},
  keywordstyle=\color{kw}\bfseries,
  commentstyle=\color{cmt}\itshape,
  stringstyle=\color{str},
  showstringspaces=false,
  columns=fullflexible,
  keepspaces=true,
  breaklines=true,
  breakatwhitespace=true,
  numbers=none,
  escapeinside={(*@}{@*)},
  aboveskip=2pt,
  belowskip=2pt,
}

\title{STILL: Recovering Lowered \textsc{Stl} Semantics for LLM-assisted \textsc{C}++ Decompilation}
\author{
  Xiaohan Wang \\
  Vanderbilt University
  \And
  Kevin Leach\thanks{\raggedright Corresponding author: \texttt{kevin.leach@vanderbilt.edu}.} \\
  Vanderbilt University
}

\begin{document}

\maketitle

\begin{abstract}
LLM-assisted decompilation improves readability and re-executability, but still underperforms on stripped C++ functions that use the Standard Template Library (STL).
Compilation, optimization, and symbol stripping remove or obscure source-level semantics such as container types and library-call structure, while traditional decompiler output often fails to recover them.
We present \textsc{STILL}, a structured semantic interface that predicts function-level STL container semantics from stripped control-flow graphs and renders them as compact hints for LLM refinement.
On \textsc{StlBench}, \textsc{STILL} predicts common container-level STL semantics, with the strongest cross-dataset results for stable \texttt{string} and \texttt{vector} slices.
On stripped HumanEval decompilation, these hints enable DeepSeek-chat refinement to reach 28.4\% executability, compared with 17.4\% for no-hint refinement and 8.9\% for raw Ghidra decompilation; hint utility is downstream-backbone-dependent, 
with decompilation-specialized models requiring lightweight adaptation to benefit from the same interface.
\end{abstract}

\section{Introduction}
\label{sec:intro}

\begin{figure}[t]
\centering
\scriptsize

% ---------- LEFT column: (a) GT source  ->  (c) Baseline output ----------
\begin{minipage}[t]{0.41\linewidth}
{\fontsize{9pt}{9.5pt}\selectfont\textbf{(a) Ground-truth C++}\par}
\begin{lstlisting}[style=cppcase, basicstyle=\ttfamily\fontsize{7.5pt}{8pt}\selectfont, breakatwhitespace=false, frame=single, framesep=3pt]
int func0(
    vector<int> arr){
  int max = -1;
  for (int i=0;
       i<arr.size(); i++)
    if (arr[i] <= i) max = i;
  return max;
}
\end{lstlisting}

\vspace{3pt}
{\fontsize{9pt}{9.5pt}\selectfont\textbf{(c) w/o STL semantics}\par}
\begin{lstlisting}[style=cppcase, basicstyle=\ttfamily\fontsize{7.5pt}{8pt}\selectfont, breakatwhitespace=false, frame=single, framesep=2pt, backgroundcolor=\color{badbg}]
int func0(vector_int_t *v) {
  int max = -1;
  for (int i=0; i<FUN_00101894(v); i++) {
    if (FUN_001018bc(v, i)->value <= i) {
      max = i;
    }
  }
  return max;
}
\end{lstlisting}
\end{minipage}\hfill
% ---------- RIGHT column: (b) Stripped pseudo  ->  (d) +hint+LoRA output ----------
\begin{minipage}[t]{0.53\linewidth}
{\fontsize{9pt}{9.5pt}\selectfont\textbf{(b) Decompiled code}\par}
\begin{lstlisting}[style=cppcase, basicstyle=\ttfamily\fontsize{7.5pt}{8pt}\selectfont, breakatwhitespace=false, frame=single, framesep=3pt]
int FUN_00101249(undefined8 v){
  int *p; ulong n;
  int ans = -1;
  int i = 0;
  while (true) {
    n = FUN_00101894(v);
    if (n <= (ulong)i) break;
    p = (int *)FUN_001018bc(v, i);
    if (*p <= i) ans = i;
    i = i + 1;
  }
  return ans;
}
\end{lstlisting}

\vspace{3pt}
{\fontsize{9pt}{9.5pt}\selectfont\textbf{(d) w/ STL semantics}\par}
\begin{lstlisting}[style=cppcase, basicstyle=\ttfamily\fontsize{7.5pt}{8pt}\selectfont, breakatwhitespace=false, frame=single, framesep=2pt, backgroundcolor=\color{goodbg}]
int func0(const vector<int>& a) {
  int ans = -1;
  for (int i=0; i<a.size(); ++i) {
    if (a[i] <= i) ans = i;
  }
  return ans;
}
\end{lstlisting}
\vspace{12pt}
\end{minipage}

\caption{\textbf{Motivating example.}
A HumanEval C++ function exposes source-level STL semantics through a \texttt{vector<int>} interface in (a).
After compilation, optimization, and stripping, the raw Ghidra decompilation in (b) fails to recover this abstraction and instead exposes low-level artifacts.
A baseline LLM refiner in (c) preserves these artifacts and fails execution, while hint-guided refinement in (d) recovers a container-level interface and passes execution tests.
}
\label{fig:motivation}
\end{figure}

Decompilation is a widely employed technique for recovering source code from stripped binaries~\cite{cifuentes1994reverse, schwartz2013native, yakdan2015nomoregotos}.  
Typically, decompilation focuses on recovering C source code to improve readability over raw assembly or machine code, thereby facilitating reverse engineering, malware analysis, and forensic investigations. 
However, C++ functions that use container types such as \texttt{vector<int>} expose rich source-level semantics that are lost during compilation, optimization, and stripping; after decompilation, these functions are often reduced to complex and unreadable pointer arithmetic, opaque data types, and unclear function calls. 
Figure~\ref{fig:motivation} shows an example function from the HumanEval dataset: rich semantics such as \texttt{vector<int>} and \texttt{size()} are lost during decompilation and replaced with low-level artifacts such as \texttt{undefined8} and \texttt{FUN\_xxx} calls.

% In source form, a short C++ function using \texttt{vector<int>} exposes its intent through the type signature, container name, and member calls.
% After compilation, optimization, and symbol stripping, however, the Ghidra decompilation given to the LLM no longer exposes that source-level abstraction.
% A \emph{stripped binary} removes symbol names and debug metadata, so a traditional decompiler often emits C-like source whose types and high-level abstractions are missing or wrong.
% Figure~\ref{fig:motivation} traces this failure end to end on a HumanEval function.
% The source in Fig.~\ref{fig:motivation}(a) uses \texttt{vector<int>}, \texttt{size()}, and indexing;
% the Ghidra decompilation in Fig.~\ref{fig:motivation}(b) replaces the vector argument with \texttt{undefined8} and opaque \texttt{FUN\_xxx} accessor calls.

%FIXME I also rewrote the second paragraph.  Consider:
In addition to the loss of semantics, decompiled code is much less likely to be \emph{recompiled} and even less likely to be \emph{re-executable}.  
This adds complexity in reverse engineering and forensics pipelines where a user may want to edit the decompiled source and still produce an executable binary. 
This issue is further amplified when focusing on C++ programs that include container types from the Standard Template Library. 
For example, on stripped HumanEval-Decompile binaries, executability falls from 51.9\% on  STL-free functions to 6.2\% on STL-bearing functions, an 8.4$\times$ gap. 
Compilation, optimization, and stripping all remove two pieces of information that 
% C/C++ % maybe??
programmers frequently use:
\emph{signature information} like \texttt{vector<string>}, and \emph{container identity} like whether an object is a \texttt{vector}, \texttt{string}, or \texttt{map}.
Because this information is removed from code when lowered during compilation, decompilers struggle to recover this information, even when assisted with or augmented by LLMs. 

% This abstraction loss creates a concentrated failure mode rather than a uniform drop in decompilation-refinement quality.
% %FIXME what?  code-generation quality? or decompilation quality?
% On stripped HumanEval-Decompile, executable success $R_\text{exec}$ falls from 51.9\% on STL-free functions to 6.2\% on STL-bearing functions, an 8.4$\times$ gap.
% The failures are also structured.
% Compilation, optimization, and stripping remove or obscure two pieces of information that source programmers normally use, and a traditional decompiler often fails to recover them in its raw decompiled output:
% \emph{signature information} such as \texttt{vector<string>},
% \emph{container identity} such as whether an object is a \texttt{vector}, \texttt{string}, or \texttt{map}.
% The LLM refiner is therefore asked to write source-level C++ from an input where the semantic channel identifying the relevant STL abstraction is missing.
% % FIXME careful, the "model" is not clearly referring to something at this point.  Which model?  I know you mean  the LLM, but that will be out of the reader's working memory at this point. 

%The key observation behind this paper 
A key insight in this paper is that residual container-level STL semantics can remain partially visible within a given binary.
\emph{Semantic residue} refers to implementation traces that remain in machine code or the control-flow graph after names and template syntax disappear, including field offsets, access widths, capacity branches, scaled indexing, tree traversals, and low-level library calls that can all serve as hints that an STL container type is being used.
For example, \texttt{vector} names disappear, but begin/end/capacity layout and reallocation branches can remain; \texttt{string} names disappear, but character-buffer accesses and length/capacity checks can remain.
This residual information cannot guarantee fully-reconstructed source types, but they can indicate which STL abstractions a function \emph{likely} uses.
The challenge is to condense this scattered low-level evidence into an interface that an LLM refiner can use.

% Existing LLM decompilation and refinement systems mainly improve the LLM refinement side of this pipeline through techniques like end-to-end neural translation~\citep{armengol2024slade,tan2024llm4decompile}, executable training-set curation~\citep{exebench}, retrieval-augmented repair~\citep{shypula2026decaf}, or context-enhanced prompting~\citep{wang2025context}.
% Adjacent binary semantic analyses recover variable-level names or types, class structure, and primitive types~\citep{wang2022tiara,schwartz2018ooanalyzer,sure2025benchmark}.
% These lines of work do not directly expose container-level STL semantics as a compact interface for LLM refinement.
% Our goal is to expose a compact semantic channel for container-level STL abstractions and test whether it helps repair the failures associated with stripped STL code.

Existing work addresses adjacent parts of this pipeline: LLM-based decompilation improves source generation or repair~\citep{armengol2024slade,tan2024llm4decompile,exebench,shypula2026decaf,wang2025context}, 
while binary analysis recovers names, types, and class structure~\citep{lacomis2019dire,chen2022dirty,xie2024resym,tie2011,retypd2016,zhu2024tygr,typeforge2025,schwartz2018ooanalyzer}.
These approaches do not directly recover STL container identities or expose
them as compact semantic hints for downstream LLM refinement.

We present \textsc{STILL}, a structured semantic interface for predicting function-level STL container semantics from stripped binaries: given a stripped control-flow graph, \textsc{STILL} predicts whether the function uses supported STL containers and which containers are present, then renders these predictions as compact hints prepended to the Ghidra decompilation before LLM refinement.
We also present \textsc{StlBench}, a 14,884-record controlled corpus built from stripped CodeContests binaries for training and evaluating \textsc{STILL}.
First, stripped binaries retain enough semantic residue to predict function-level STL container semantics: \textsc{STILL} reaches 80.4\% macro-F1 on held-out CodeContests and 89.0\% macro-F1 on the stable HumanEval-C++ \texttt{string}/\texttt{vector} transfer slice.
Second, predicted semantic hints improve downstream decompilation: DeepSeek-chat reaches 28.4\% stripped HumanEval $R_\text{exec}$ with predicted semantic hints, compared with 17.4\% for no-hint refinement and 8.9\% for raw Ghidra decompilation.
Third, hint utility depends on the downstream backbone: an oracle hint is nearly inert for \textsc{LLM4Decompile-Ref-6.7B-v2}~\citep{tan2024llm4decompile} at inference time, but lightweight LoRA adaptation~\citep{hu2021lora} enables the model to benefit from the same interface and raises STL-slice $R_\text{exec}$ from 1.1\% to 26.9\%.

Our contributions are: (1) We identify STL semantic loss during compilation, optimization, and stripping as a specific failure channel in stripped C++ LLM-assisted decompilation, separating signature loss and container identity loss from generic refinement errors. (2) We introduce \textsc{STILL}, which predicts function-level STL container semantics from stripped CFGs and exposes them as compact semantic hints. \textsc{StlBench} provides supervision and evaluation for this interface. (3) We show that predicted semantic hints improve executable decompilation under a fixed chat model, while specialized decompilation backbones may require adaptation to consume the same interface. (4) We release the artifact.

\section{\textsc{StlBench}: A Controlled Corpus for STL Semantic Extraction}
\label{sec:dataset}
% First, we present a dataset called \textsc{StlBench} to provide the controlled data needed to study STL semantic loss during compilation, optimization, and stripping rather than a new general-purpose decompilation benchmark.
The previous section frames STL recovery as a semantic-interface problem;
we now describe the controlled corpus used to train and evaluate that interface.
A controlled corpus is needed
because existing decompilation resources support general binary-to-source recovery
or variable-level type recovery~\citep{exebench,tan2024llm4decompile,sure2025benchmark,zhu2024tygr,wang2022tiara},
but do not isolate how STL container information
survives stripping
and affects downstream LLM refinement.
We therefore build \textsc{StlBench},
which aligns stripped C++ functions,
raw Ghidra decompiler output,
and source-derived container labels.

%FIXME you may want one more half sentence about what design properties lead to that isolation.  The "Construction" subsection may address this already, but this could be something like.... "StlBench considers functions that use exactly 1 STL container type" or "StlBench includes C++ functions that call a constructor of a STL Container type..."  

%FIXME this feels like an okay place to put a diagram or figure describing how the dataset is constructed, labeled, and how many projects, functions, files, etc. are considered.  Not necessary unless you have time.  

\subsection{Construction}
\label{sec:dataset-source}
We start from CodeContests~\citep{li2022alphacode} C++ solutions
and select self-contained target functions:
non-\texttt{main} functions
with at least five source lines
and no calls to other user-defined functions
in the same file.
To keep the setting controlled despite imbalanced STL usage
in GitHub C++ code,
we restrict supervision to five containers
(\texttt{map}, \texttt{queue}, \texttt{set}, \texttt{string}, and \texttt{vector})
and use per-class caps with explicit with-STL/no-STL quotas.
The resulting corpus is roughly 60\% with-STL
and 40\% no-STL,
preserving no-STL functions as a control group.
% We exclude \texttt{std::array} because it typically compiles to ordinary stack memory and leaves no reliable container-specific signal.
%FIXME I think this weakens the paper substantially as phrased ---- you may want another clause here saying that this is okay because it is effectively indistinguishable from a  standard stack-based array. 
We exclude \texttt{std::array}
because it is usually lowered to fixed-size stack or inline storage,
making its compiled representation effectively indistinguishable
from an ordinary C-style array;
unlike heap-backed or node-based STL containers,
it does not preserve a distinct container-level runtime abstraction.
Each selected function is compiled with \texttt{g++~-g}
using O0--O3;
the debug flag is used only to locate the target function
and derive supervision,
while model inputs are extracted from stripped worker copies.
The release contains 3{,}894 unique source functions
and 14{,}884 optimization-level records
after compilation, stripping, disassembly,
CFG extraction,
and label-alignment failures are removed
(81\% retention).

\subsection{Semantic Supervision}
% Function-level labels define which source abstractions should be recovered from the stripped binary after compilation, optimization, and stripping obscure them.
% %FIXME how did you do the labeling? Manually?  Is there an appendix or something that shows how you did it?  Also this is passive voice.  consider instead...
% %   We manually labeled each function according to whether abstractions described in the source code should be recoverable from the stripped binary.
% % FIXME ^ but this invites a new question -- what basis do you use to determine whether it "should" be recovered?   This is one of those places where I would expect to see something like "For each sample, we label is a "recoverable" when that sample contains source code with an unambiguous reference to one of our 5 STL types, contains XYZ basic blocks..."    Remember, the reader has to be able to reproduce your results, including the labeling you made for this dataset. 
% We derive multi-label annotations by intersecting canonical \texttt{std::} template instantiations with AST-level usage inside the selected function body, filtering comments, string literals, and unused declarations.
% For nested declarations, only the outermost container is labeled, matching the container-level storage and indexing logic that can survive compilation.
% These labels supervise the binary STL gate and function-level container head.
% The source-derived 
% sidecar  %FIXME sidecar??  I don't quite understand what this means here. 
% provides supervision but is never concatenated into the input features.

\textsc{StlBench} uses source-derived labels
to define the function-level prediction target for \textsc{STILL}.
Each label records whether the selected source function
uses one of the five supported STL containers:
\texttt{map}, \texttt{queue}, \texttt{set}, \texttt{string}, or \texttt{vector}.
These labels are container-level targets,
not claims that exact source-level C++ types
can be reconstructed from the binary.

We generate these labels automatically
by analyzing the original source code.
Before labeling,
we remove comments and string or character literals,
then detect top-level container uses
in the function signature,
direct declarations inside the target function,
alias-expanded declarations,
macro-expanded container forms,
and global container variables
used by the target function.

The labeling procedure is multi-label,
so a function may contain multiple supported containers.
To avoid treating implementation details
as independent supervision targets,
we count only top-level container occurrences.
The resulting source-derived supervision metadata
is used only for training and evaluation;
it is never concatenated into the stripped CFG features
or the raw decompiler output seen by the LLM refiner.

\subsection{Controls and Release}
% We split % FIXME split into what?
% by CodeContests problem identifier rather than function identifier and remove near-duplicates against HumanEval and decomp-eval using binary-embedding cosine similarity $>0.95$.
During dataset construction,
we keep at most one target function
per CodeContests problem
to reduce overlap between functions
from the same programming task.
We then split train, validation, and test partitions
by target function,
so optimization-level variants of the same function
cannot cross split boundaries.
We apply two additional leakage checks.
First,
we remove near-duplicates against HumanEval
and decomp-eval
using binary-embedding cosine similarity above $0.95$.
Second,
we audit residual external-call names
in stripped assembly
and find that direct container-name cues
are rare overall
(Appendix~\ref{app:symbol-audit});
the task is therefore not reducible
to surviving STL symbols.
\textsc{StlBench} covers five common containers---\texttt{map}, \texttt{queue},
\texttt{set}, \texttt{string}, and \texttt{vector}---that span contiguous,
string-buffer, ordered-associative, and adaptor-like implementation families.
Its label space targets top-level uses of these containers in stripped
\texttt{g++}/x86-64/libstdc++ binaries; nested container composition,
iterator-level behavior, and \texttt{std::array} are outside this scope.
We release the g++/x86-64/libstdc++ instantiation
used in this paper,
including stripped CFGs,
raw Ghidra decompiler output,
function-level labels,
and the construction pipeline
needed to regenerate variants
under other toolchains or architectures.

\section{Approach}
\label{sec:approach}

%[x]FIXME I always advise ensuring that you never have "orphaned section headers" where you have a header followed immediately by another header.   
% Consider adding something brief here like "In this section, we present STILL.  We describe the X, Y, and Z parts of STILL.

This section presents \textsc{STILL}'s semantic interface, explains why STL container residues remain recoverable from stripped binaries, describes the typed CFG extractor, and shows how the recovered interface conditions LLM-based decompilation.

\subsection{Semantic Interface}
\label{sec:semantic-interface}

\begin{figure*}[!t]
    \centering
    \includegraphics[width=\linewidth,height=\textheight,keepaspectratio]{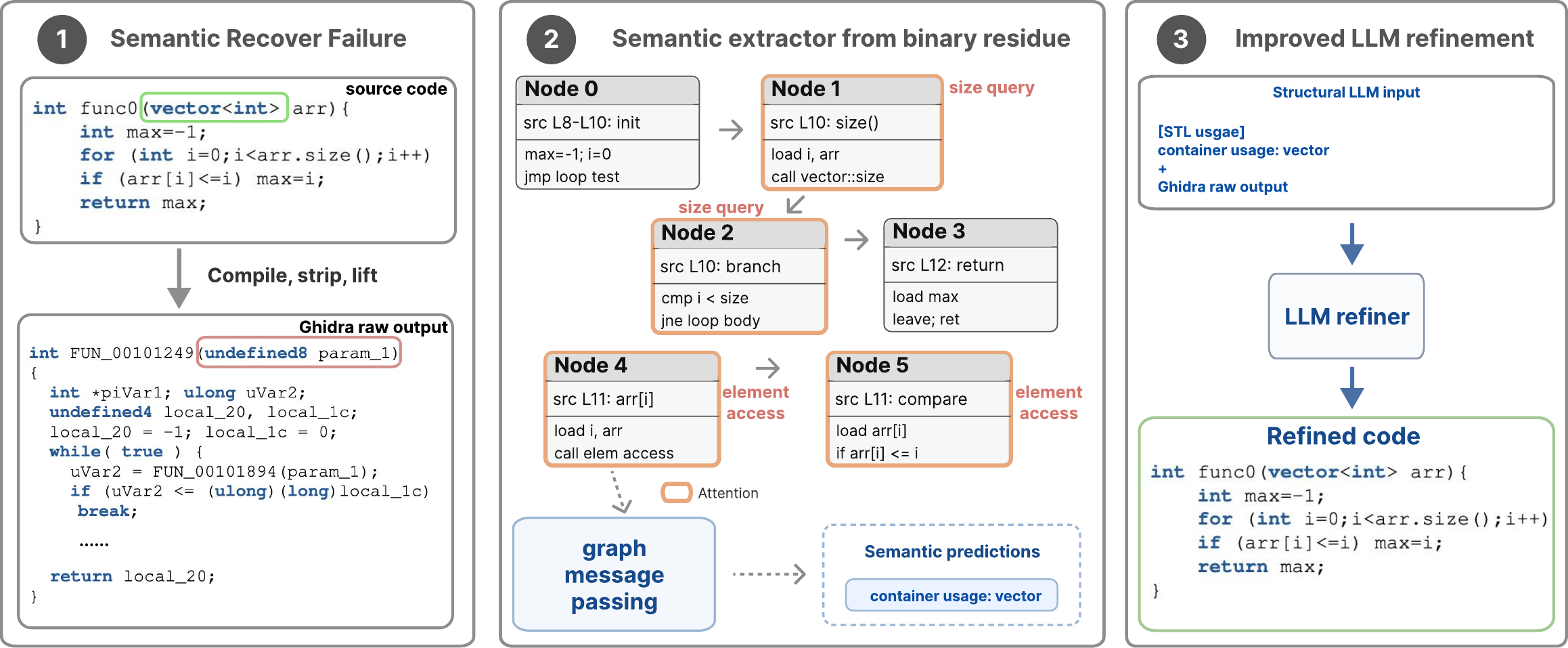}
    \caption{STILL predicts STL semantics from binary residues for LLM refinement.}
    \label{fig:approach-overview}
\end{figure*}

\autoref{fig:approach-overview} summarizes the \textsc{STILL} pipeline.
\textsc{STILL} predicts the STL semantics that compilation, optimization, and stripping obscure and that a traditional decompiler often fails to recover before LLM refinement.
Given a stripped C++ function, we first extract a typed control-flow graph $G=(V,E,X,R)$, where $V$ is the set of basic blocks, $E$ contains control-flow edges, $X$ is the per-block feature matrix, and $R:E\rightarrow\{1,\ldots,K\}$ assigns each edge one of $K{=}6$ relations: true branch, false branch, unconditional jump, fallthrough, call, or return.
The extractor maps this graph to two outputs.
The binary STL gate $\hat{g}\in[0,1]$ predicts whether the function uses any supported STL container at all.
The container vector $\hat{y}\in[0,1]^C$ is a function-level multi-label prediction over \texttt{map}, \texttt{queue}, \texttt{set}, \texttt{string}, and \texttt{vector}.

These outputs deliberately form a compact interface rather than a reconstructed source program.
The gate prevents unsupported or no-STL functions from receiving spurious hints.
The container vector restores the source-level abstraction that would normally appear in type signatures and object declarations.
At inference time, \textsc{STILL} renders $(\hat{g},\hat{y})$ as a 
structured hint %FIXME you may want an example of what this looks like
and supplies it together with Ghidra decompilation
%FIXME as with previous comments
to the LLM refiner; the downstream objective remains executable recovery, measured by $R_\text{exec}$, the fraction of functions whose decompiled source compiles and passes the original tests.  %[x]FIXME I know you refer to this elsewhere, but it may be worth defining again here since this is the first reference to it in this section.

\subsection{Why STL Residues Are Recoverable}
\label{sec:binary-residue}

Compilation removes names and template syntax, but it does not remove the implementation 
idiosyncrasies %[x]FIXME hmmm... "regularities" ?   maybe "idiosyncrasies" is better?
that make STL containers work.
For STL containers, the useful traces are lossy regularities rather than reliable fingerprints: \texttt{vector} often leaves scaled pointer indexing and begin/end/capacity-like offsets, \texttt{string} leaves byte-buffer accesses and length or capacity checks, and ordered containers leave comparator branches and pointer chasing through tree-shaped nodes.
These cues explain why \texttt{string} and \texttt{vector} are usually recovered more accurately than \texttt{map} and \texttt{set}, whose key-value and key-only variants often lower to similar traversal code.  %[x]FIXME consider avoiding weasely words like this.  Presumably you mean that string and vector are more accurately recovered; the problem is still not "easy".

The extractor therefore treats residue evidence conservatively.
It recovers a missing container-identity channel for the LLM, but it does not claim to reconstruct full source types or disambiguate every STL implementation detail.

\subsection{Extractor Implementation}
\label{sec:extractor-implementation}

The extractor implements the interface above with a typed CFG encoder.
The method's main contribution is the semantic interface---its function-level label definition, compact hint format (e.g., \texttt{// STL containers: vector}), and insertion before the Ghidra decompilation in the LLM prompt.  The RGCN below is one implementation for predicting these labels from CFG residue; it is not a new GNN architecture or a separate contribution.
We use angr's \texttt{CFGFast} analysis~\citep{shoshitaishvili2016angr} to recover basic blocks and relation-typed control-flow edges from the stripped binary.
Ghidra is used separately as the decompiler frontend because it is a mature open-source reverse-engineering framework with a widely used decompiler, and its output forms the baseline decompilation context for LLM refinement.
For the semantic extractor, however, we use angr because CFG recovery is a first-class analysis in angr, and \texttt{CFGFast} directly exposes a scriptable architecture-level graph over binary basic blocks.
The choice of angr is therefore replaceable: any analysis backend that provides equivalent basic blocks, typed control-flow edges, and block-local residue features could be substituted.
%[x]FIXME I think you'll get questions about why you're using angr for the CFG when Ghidra has CFG extraction as well.  If you're using Ghidra for the "pseudo code" then why not Ghidra for the cfg as well?  Or vice versa with angr?
For the selected extractor, each basic block $v$ is represented by STL-specific residue features, written compactly as
$\mathbf{x}_v=[\textsc{ContRes}(v)\Vert\textsc{TreeRes}(v)]$.
For each basic block, \textsc{STILL} computes residue features from stripped assembly only,
not from source names or decompiler output.
The node representation concatenates two feature blocks:
a 33-dimensional \textsc{ContRes} vector for contiguous-container residue,
and a 68-dimensional \textsc{TreeRes} vector for ordered-container residue.
\textsc{ContRes} summarizes layout offsets, access widths, stride/indexing cues,
and capacity, copy, or mutation predicates.
\textsc{TreeRes} summarizes tree-node layout, compare/branch and pointer-chasing cues,
map/set payload evidence, and CFG-region context.
Count features use $\log(1+x)$ scaling,
while structural cues are binary indicators or normalized scores.
Appendix~\ref{app:extractor-implementation} gives the feature-level implementation details.
The architectural novelty is not the RGCN itself,
but the STL-specific residue representation and the semantic interface that connects recovered facts to LLM refinement.

Node features are passed through a 3-layer RGCN with relation-specific weight matrices for the six CFG edge types, residual connections, and LayerNorm.
The resulting node embeddings $\mathbf{h}_v$ feed the two interface heads.
A linear gate head over a mean-pooled graph representation predicts $\hat{g}$ and suppresses container hints when the score falls below $0.5$.
A per-class attention pooling head builds a class-conditioned graph vector for each container and predicts the multi-label container vector $\hat{y}$.
All heads are trained jointly with class-weighted binary cross-entropy, $\mathcal{L}=\mathcal{L}_\text{type}+\lambda_\text{gate}\mathcal{L}_\text{gate}$.

\subsection{Hint-Conditioned Decompilation}
\label{sec:hint-llm}

The recovered interface is rendered as a compact container list, for example that a function likely uses \texttt{string} and \texttt{vector}.
This rendering tests whether decompilation gains come from restoring the missing container identity channel without requiring a reconstructed source program.
The full prompt template is included in the appendix.

For chat-model refinement, the hint is prepended to or inserted into the Ghidra decompilation and the model is asked to emit compilable C++ source.
For the specialized LLM4Decompile-Ref-6.7B-v2 backbone, we additionally evaluate hint-aware LoRA adaptation.
This adaptation is not a separate contribution.
It is an evaluation condition used to test whether a specialized downstream decompilation backbone can learn to consume the same semantic interface.
The adapter uses rank $r{=}16$, $\alpha{=}32$, all linear targets, learning rate $1{\times}10^{-4}$, three epochs, and cutoff length $4096$; it is trained on \textsc{StlBench} pseudo-to-source pairs with oracle hints, then evaluated with GNN-recovered hints in the deployment-realistic setting.

\section{Evaluation}
\label{sec:eval}

%Our evaluation traces the container hint from recovery to deployment via three research questions:

We answer three research questions:
\smallskip

\noindent\textbf{RQ1.} Are missing source-level STL semantics recoverable from
stripped binaries, and when is recovery reliable or ambiguous?

\noindent\textbf{RQ2.} Do recovered container semantics improve both
executability and readability?

\noindent\textbf{RQ3.} When does a downstream refinement model consume
recovered semantics?
%[x]FIXME careful with this last question.  The "decompiler" could refer to Ghidra.  But presumably you mean the downstream LLM refinement model. 

\subsection{Experimental Setup}
\label{sec:setup}

\paragraph{Datasets.}
We train the semantic extractor on \textsc{StlBench} (Sec.~\ref{sec:dataset}),
which contains 3{,}894 CodeContests functions compiled at four optimization levels,
yielding 14{,}884 stripped opt-level records with function-level labels over
\texttt{map}, \texttt{queue}, \texttt{set}, \texttt{string}, and \texttt{vector}.
RQ1 evaluates both the CodeContests held-out split ($\approx$750 graphs per opt level)
and zero-shot transfer to stripped HumanEval-C++
(643 retained opt-level records from 164 functions $\times$ 4 optimization levels).
Because HumanEval-C++ has broad support only for \texttt{string} and \texttt{vector}
in our five-class label space, we report cross-dataset RQ1 under the stable
\texttt{string}/\texttt{vector} projection rather than letting absent labels dominate macro-F1.
RQ2 and RQ3 use stripped HumanEval-C++ to evaluate downstream decompilation.

\paragraph{Architecture.}
%FIXME I am slightly confused by the phrasing of "continuous" and "tree" here.  What is continuous here? and what tree is being used?
Unless otherwise stated, \textsc{STILL} uses the selected continuous+tree residue representation:
continuous layout/access/width/stride features together with cross-basic-block tree-region features.
A 3-layer RGCN over typed CFG edges aggregates these node features with per-class attention pooling.
We report per-opt/per-class breakdowns in Appendix~\ref{app:rq1-perclass}.

\paragraph{Metrics.}
For RQ1, each function is a multi-label prediction over the five supported STL container classes;
we report per-class F1, macro-F1, and gate F1 for detecting whether any supported STL container is present.
For RQ2/RQ3, we report compile rate $R_\text{comp}$ and execution success rate $R_\text{exec}$,
where $R_\text{exec}$ requires the generated C++ to compile and pass the original HumanEval unit tests.
We use edit similarity and paired LLM-judge readability as secondary output-quality measures.

\paragraph{Baselines and comparisons.}
RQ1 baselines rule out three shortcuts:
a majority-class prior tests label imbalance,
a Ghidra regex readout tests symbol or Ghidra-decompilation leakage,
and a zero-shot LLM classifier tests prompt-only guessing from Ghidra decompilation.
We do not include TIARA~\citep{wang2022tiara} as a direct baseline because it requires a known variable address;
our setting starts from the CFG of target function and predicts a function-level container labels without requiring address-anchored variables.
RQ2 compares raw Ghidra decompilation and no-hint LLM refinement against the same Ghidra decompilation with a compact recovered-container hint.
RQ3 compares the specialized downstream decompilation backbone with and without hint-aware adaptation to test whether the model can consume the semantic interface.

\begin{figure}[!t]
\centering
\scriptsize
\begin{minipage}[t]{0.48\linewidth}
\centering{\fontsize{9pt}{9.5pt}\selectfont\textbf{DeepSeek refinement}\par}
\begin{lstlisting}[style=cppcase, basicstyle=\ttfamily\fontsize{7.5pt}{8pt}\selectfont, breakatwhitespace=false, frame=single, framesep=2pt, backgroundcolor=\color{badbg}]
struct Vector {
int *data,*end,
    *cap; };

Vector* func0(
Vector* vec,
int start) {
/* init omitted */
int count=1;
while(count!=
      start) {
int* e=vec->end;
int next=
  *(e-1)+2;
if(e==vec->cap)
  grow_vector(
    vec,e,&next);
else {
  *e=next;
  vec->end=e+1; }
count++; }
return vec; }
\end{lstlisting}
\vspace{-2pt}
\end{minipage}\hfill
\begin{minipage}[t]{0.48\linewidth}
\centering{\fontsize{9pt}{9.5pt}\selectfont\textbf{DeepSeek + Type Hint}\par}
\begin{lstlisting}[style=cppcase, basicstyle=\ttfamily\fontsize{7.5pt}{8pt}\selectfont, breakatwhitespace=false, frame=single, framesep=2pt, backgroundcolor=\color{goodbg}]
std::vector<int>&
func0(
std::vector<int>&
vec,
int start) {
vec.clear();
vec.reserve(1);
vec.push_back(
  start);

int i=1;
if(1<start) do {
int next=
  vec.back()+2;
vec.push_back(
  next);
i++;
} while(start!=i);
return vec; }
\end{lstlisting}
\vspace{-2pt}
\end{minipage}
\caption{\textbf{Qualitative example.}
On the same failed HumanEval pair (\texttt{100\_O1}), the type hint does not make the program fully executable, but it replaces a synthetic pointer-level \texttt{Vector} reconstruction with the source-level \texttt{std::vector<int>} interface.
}
\label{fig:rq2-quality-example}
\end{figure}

\subsection{RQ1: Recoverability of Missing STL Semantics}
\label{sec:rq1}

\begin{table}[!t]
    \centering
    \small
    \caption{Function-level STL semantic extraction. CC denotes the CodeContests held-out split; HE denotes the HumanEval-C++ subset.}
    \label{tab:rq1-semantic-extraction}
    \setlength{\tabcolsep}{0pt}
    \renewcommand{\arraystretch}{1.10}
    \begin{tabular*}{\linewidth}{@{\extracolsep{\fill}}llrrrrr@{\hspace{5pt}}rr@{}}
        \toprule
        & 
        & \multicolumn{5}{c}{Container F1} 
        & \multicolumn{2}{c}{Summary} \\
        \cmidrule(lr){3-7}
        \cmidrule(l){8-9}
        Data & Method 
        & map & queue & set & string & vector 
        & M-F1 & Gate \\
        \midrule
        CC & Majority 
        & 0.0 & 0.0 & 0.0 & 0.0 & 50.7 
        & 10.1 & 74.8 \\

        CC & Ghidra readout 
        & 1.0 & 0.0 & 1.3 & 76.3 & 30.2 
        & 21.7 & 50.4 \\

        CC & Permuted 
        & 56.8 & 35.9 & 45.5 & 50.0 & 68.3 
        & 50.1 & 90.7 \\

        CC & \textbf{\textsc{Ours}} 
        & \textbf{78.3} & \textbf{87.4} & \textbf{66.9} 
        & \textbf{88.0} & \textbf{81.6} 
        & \textbf{80.4} & \textbf{95.4} \\
        \midrule
        HE & LLM zero-shot 
        & -- & -- & -- & 31.6 & 22.2 
        & 26.9 & 32.8 \\

        HE & \textbf{\textsc{Ours}} 
        & -- & -- & -- & \textbf{91.8} & \textbf{86.4} 
        & \textbf{89.0} & \textbf{98.6} \\
        \bottomrule
    \end{tabular*}
\end{table}

We ask whether our approach recovers function-level STL container semantics from stripped binaries.
We evaluate in-domain on CodeContests (E1) and zero-shot on HumanEval-C++ (E2);
the full setup is described in Sec.~\ref{sec:setup}.

The first question is whether a stripped binary still contains enough evidence to identify which STL containers a function uses.
On the CodeContests held-out split,
\textsc{STILL} predicts the five container classes with 80.4\% macro-F1 and 95.4\% gate F1 (Table~\ref{tab:rq1-semantic-extraction}).
Without any HumanEval supervision,
the same extractor reaches 89.0\% macro-F1 on the HumanEval \texttt{string}/\texttt{vector} slice,
showing that the strongest container residues transfer across datasets.

This result is not explained by straightforward shortcuts.
The majority-class prior is a label-imbalance control: it ignores the binary and predicts from the training-label distribution alone.
It reaches only 10.1\% macro-F1, so the recovered labels cannot be explained by class frequency.
The Ghidra readout is a leakage control: it searches stripped Ghidra decompilation and residual library-call strings for explicit container-name cues.
It reaches only 21.7\% macro-F1, and direct container-name cues appear in only 218 of 8{,}900 STL-bearing records (2.4\%),
mostly for \texttt{string} (Appendix~\ref{app:symbol-audit}),
showing that Ghidra does not already expose most STL semantics as text.
The zero-shot LLM baseline is a prompt-only control: it asks a general code model to infer container labels directly from Ghidra decompilation without training our extractor.
On HumanEval, this baseline reaches only 26.9\% macro-F1 on the same stable projection,
so the transfer result is not prompt-only guessing from Ghidra decompilation.

%We also avoid an apples-to-oranges comparison with TIARA~\citep{wang2022tiara}.
We also ensure an apples-to-apples comparison with TIARA~\citep{wang2022tiara}.
TIARA is the closest prior system for C++ container-type recovery, but it assumes a known variable address and predicts variable-level labels.
Our RQ1 setting provides no address-anchored variable query: the input is a whole stripped function CFG, and the output is the function-level set of STL containers to be passed downstream as a compact semantic hint.

%The controls show why residue alignment matters.
The selected continuous+tree residue model reaches 80.4\% macro-F1 and 95.4\% gate F1 because its features explicitly encode layout/access and tree-region evidence.
When the continuous residue features are permuted across graphs with similar node counts, per-opt macro-F1 drops to 42.0/53.0/56.0/55.1 for O0--O3.
This sanity control preserves the extra feature dimensions and model plumbing, but breaks the alignment between residue and function semantics; the drop indicates that recoverability comes from semantically aligned binary residue rather than from capacity alone.

\label{sec:rq1-perclass}
Per-class F1 (Table~\ref{tab:rq1-semantic-extraction}) splits the five classes into three tiers driven by what survives stripping:
\texttt{queue} and \texttt{string} retain distinctive call sites and field layouts and are recovered reliably;
\texttt{vector} keeps its begin/end/capacity triple but shares this signature with non-STL pointer triplets;
\texttt{map} and \texttt{set} both lower to the same libstdc++ \texttt{\_Rb\_tree\_node} layout and lose the comparator/allocator tag that separates them.

We also evaluate the frozen extractor and downstream hint protocol on
a fresh synthetic C++17 set constructed locally; Appendix~\ref{app:contamination-control}
reports the contamination-control protocol and results.

% \begin{tcolorbox}[colback=gray!5,colframe=gray!40!black,boxrule=0.4pt,boxsep=-1mm]
% \textbf{Answer to RQ1.}
% Missing STL semantics are recoverable through type-specific binary residues rather than a single generic signal:
% contiguous containers expose layout/access residues, such as \texttt{vector} begin/end field loads, pointer-difference size computation, and scaled indexed loads over adjacent elements (Fig.~\ref{fig:vector-residue-case});
% associative containers require cross-basic-block tree-region evidence,
% and adaptors expose a mixture of wrapper and backing-container residues.
% \end{tcolorbox}

\subsection{RQ2: Do Recovered Semantics Improve Decompilation Quality?}
\label{sec:rq2}

We evaluate whether recovered container semantics improve decompilation quality along two important axes:
\emph{executability}, where the refined program must compile and pass tests,
and \emph{readability}, where the refined program should expose source-level C++ abstractions rather than decompiler artifacts.
The experiment compares paired refinements of the same stripped Ghidra decompilation with and without a recovered container-type hint.

\paragraph{Executability.}
We compare two conditions on stripped HumanEval ($n{=}598$,
restricted to functions with non-empty Ghidra decompilation;
DeepSeek-chat,
no fine-tuning).
The \textsc{ghidra} row is the raw Ghidra decompilation submitted directly without any LLM refinement,
which exposes the floor imposed by Ghidra's \texttt{undefined8}/\texttt{ulong} type aliases.
The \textsc{+type semantic} control keeps the Ghidra decompilation unchanged but adds a separate prompt section containing only the recovered STL container list,
without line anchors or additional labels.
Against the no-hint zero-shot refinement, this compact semantic channel raises overall $R_\text{exec}$ from $17.4\%$ to $28.4\%$ and HasSTL $R_\text{exec}$ from $5.5\%$ to $18.6\%$ (Table~\ref{tab:rq2-hint-format}).
Raw Ghidra provides the lower bound: it reaches $0.0\%$ HasSTL $R_\text{exec}$ but $49.1\%$ NoSTL $R_\text{exec}$,
confirming that STL-bearing functions are the hard slice rather than merely a random subset of HumanEval.
NoSTL functions already compile and run at much higher rates across all prompt conditions,
so the large HasSTL gain is the main executability evidence for the recovered semantic hint.

\begin{table}[!t]
    \centering
    \small
    \caption{Semantic-interface ablation for \textsc{STILL} on stripped HumanEval ($n{=}598$). Entries are percentages.}
    \label{tab:rq2-hint-format}
    \setlength{\tabcolsep}{0pt}
    \renewcommand{\arraystretch}{1.12}
    \begin{tabular*}{\linewidth}{@{\extracolsep{\fill}}lrrrrrr@{}}
        \toprule
        & \multicolumn{2}{c}{Overall}
        & \multicolumn{2}{c}{HasSTL}
        & \multicolumn{2}{c}{NoSTL} \\
        \cmidrule(lr){2-3}
        \cmidrule(lr){4-5}
        \cmidrule(l){6-7}
        Interface 
        & $R_c$ & $R_x$ 
        & $R_c$ & $R_x$ 
        & $R_c$ & $R_x$ \\
        \midrule
        Raw Ghidra 
        & 8.9 & 8.9 & 0.0 & 0.0 & 49.1 & 49.1 \\
        \midrule

        \multicolumn{7}{@{}l}{\textit{DeepSeek V4.0}} \\
        Zero-shot
        & 21.9 & 17.4 & 8.0 & 5.5 & 85.2 & 71.3 \\
        +Type Semantic
        & 35.8 & 28.4 & 24.5 & 18.6 & 87.0 & 73.1 \\
        $\Delta$
        & +13.9 & +11.0 & +16.5 & +13.1 & +1.8 & +1.8 \\

        \bottomrule
    \end{tabular*}
\end{table}

\begin{figure}[!t]
    \centering
    \includegraphics[width=0.95\linewidth]{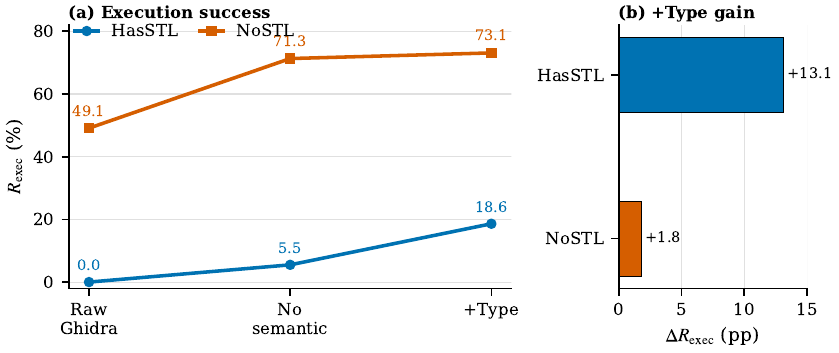}
    \caption{Type hints improve execution mainly on STL-bearing functions, with much smaller gains on NoSTL functions.}
    \label{fig:rq2-split-gain}
\end{figure}

\begin{table}[!t]
    \centering
    \small
    \caption{Paired readability of DeepSeek refinements. RelRead $>0$ favors
    \textsc{+type semantic}.}
    \label{tab:rq2-readability}
    \setlength{\tabcolsep}{0pt}
    \renewcommand{\arraystretch}{1.08}
    \begin{tabular*}{\linewidth}{@{\extracolsep{\fill}}lrrrr@{}}
        \toprule
        Cohort & $\Delta R_x$ (pp) & RelRead $\uparrow$ & $\Delta$STL $\uparrow$ & $\Delta$Art. $\downarrow$ \\
        \midrule
        All & +11.2 & +0.338 & +0.890 & -0.945 \\
        HasSTL & +13.3 & +0.354 & +1.082 & -1.173 \\
        Both fail $R_x$ & 0.0 & +0.184 & +1.167 & -1.300 \\
        \bottomrule
    \end{tabular*}
\end{table}

\paragraph{Readability.}
%FIXME I encourage you to find citations to support the validity of this methodology.  We need evidence that this sort of LLM judgment is sufficiently convincing to the community to apply here. 
%reply: I am using the concept of R2I; 
%FIXME Also, does it matter if you're using DeepSeek both for the decompilation and for the readability measure?  Or should a different model have been used?
Execution success is a strict semantic endpoint, but it misses cases where a decompiler has moved from low-level binary residue toward readable source-level structure.
We therefore define \emph{Relative Readability} (RelRead) as a paired LLM-as-a-judge score.
For each sample, a DeepSeek judge sees the same Ghidra decompilation, refinement A without semantic hints, refinement B with the \textsc{+type semantic} hint, and the recovered type hint supplied to B.
The judge is instructed to score relative human readability rather than functional correctness:
positive scores mean B is more readable, negative scores mean A is more readable, and zero means no meaningful difference.
The prompt rewards idiomatic STL abstractions, clearer identifiers and control flow, and fewer raw pointer or memory-layout artifacts, while penalizing misleading abstractions and decompiler residues such as \texttt{undefined*}, \texttt{FUN\_*}, \texttt{DAT\_*}, casts, and raw address arithmetic.

In paired readability comparisons, \textsc{+type semantic} wins 305, ties 130,
and loses 162, with a mean RelRead of $+0.34$
(Table~\ref{tab:rq2-readability}).
We also report two transparent proxies for the same readability claim:
the change in source-level STL abstraction mentions and the change in low-level decompiler artifacts.
On all pairs, the hinted refinement adds $+0.89$ STL abstractions on average and removes $0.95$ artifacts.
The key control is the both-fail subset:
among 414 pairs where both variants still fail $R_\text{exec}$, \textsc{+type semantic} remains more readable on average (RelRead $+0.18$), introduces more STL-level abstractions ($+1.17$), and removes more decompiler artifacts ($-1.30$).
Figure~\ref{fig:rq2-quality-example} shows a representative both-fail case: the no-hint output reconstructs a hand-written \texttt{Vector} with manual reallocation, while the hinted output recovers the source-level \texttt{std::vector<int>} interface.

% \begin{tcolorbox}[colback=gray!5,colframe=gray!40!black,boxrule=0.4pt,boxsep=-1mm]
% \textbf{Answer to RQ2.}
% Recovered container semantics improve stripped C++ decompilation.
% A compact \textsc{+type semantic} hint raises HasSTL $R_\text{exec}$ from 5.5\% to 18.6\% and improves paired readability even when both variants fail the tests, indicating that container identity helps restore source-level STL abstractions before full functional correctness is achieved.
% \end{tcolorbox}

%FIXME As above, careful with the phrasing here.  The "decompiler" is Ghidra.  You are improving the decompilation by feeding things through an LLM.  
\subsection{RQ3: When Does a Downstream Refinement Model Consume Hints?}
\label{sec:rq3-downstream}

We next isolate whether the specialized downstream decompilation backbone can use STL hints without adaptation.
At fixed weights, an oracle STL hint changes $R_\text{exec}$ by only $+0.2$\,pp, showing that the prompt-side fact is not enough for this downstream model.
LoRA fine-tuning on the same hint-augmented format unlocks the signal: with oracle hints, $R_\text{exec}$ rises from 11.8\% to 30.8\% overall and from 1.1\% to 26.9\% on STL-bearing functions (Table~\ref{tab:rq3-sft}). Replacing oracle hints with GNN-recovered hints retains 21.0/26.9 of the STL-slice oracle gain, or ${\sim}$78\%.
Adding an oracle hint to the no-hint-trained adapter changes STL $R_\text{exec}$ by only $+0.4$\,pp, whereas the hint-aware adapter gains $+18.1$\,pp (Table~\ref{tab:rq3-hint-consumption}).

\begin{table}[!t]
    \centering
    \small
    \caption{Diagnostic for whether adapter training induces hint use. Values are STL-slice $R_\text{exec}$.}
    \label{tab:rq3-hint-consumption}
    \setlength{\tabcolsep}{4pt}
    \renewcommand{\arraystretch}{1.05}
    \resizebox{\linewidth}{!}{%
    \begin{tabular}{lrrr}
        \toprule
        Adapter training & No hint at test & Oracle hint at test & Change \\
        \midrule
        SFT, no hints & 23.5 & 23.9 & $+0.4$\,pp \\
        SFT, oracle hints & 8.0 & 26.1 & $+18.1$\,pp \\
        \bottomrule
    \end{tabular}%
    }
\end{table}

\begin{table}[!t]
    \caption{Controlled ablation of the SFT contribution on full stripped HumanEval-C++ ($n{=}584$).
    All rows use LLM4Decompile-Ref-6.7B-v2~\citep{tan2024llm4decompile}.
    STL is the subset with at least one STL container in ground truth ($n{=}476$).}
    \label{tab:rq3-sft}
    \centering\small
    \setlength{\tabcolsep}{4pt}
    \resizebox{\linewidth}{!}{%
    \begin{tabular}{l cc cc}
        \toprule
        & \multicolumn{2}{c}{Overall} & \multicolumn{2}{c}{Stratified $R_\text{exec}$} \\
        \cmidrule(lr){2-3}\cmidrule(lr){4-5}
        Condition & $R_\text{comp}$ & $R_\text{exec}$ & STL & NoSTL \\
        \midrule
        No-SFT, no hint                  & 16.8 & 11.6 & 1.3 & 57.4 \\
        No-SFT, oracle hint              & 18.3 & 11.8 & 1.1 & 59.3 \\
        \textbf{SFT, oracle hint}        & \textbf{64.0} & \textbf{30.8} & \textbf{26.9} & 48.1 \\
        \textbf{SFT, GNN hint}           & \textbf{51.7} & \textbf{25.7} & \textbf{21.0} & 46.3 \\
        \midrule
        $\Delta_\text{hint}$             & $+1.5$\,pp & $+0.2$\,pp & $-0.2$\,pp & $+1.9$\,pp \\
        $\Delta_\text{GNN gap}$          & $-12.3$\,pp & $-5.1$\,pp & $-5.9$\,pp & $-1.8$\,pp \\
        \bottomrule
    \end{tabular}%
    }
\end{table}

% \begin{tcolorbox}[colback=gray!5,colframe=gray!40!black,boxrule=0.4pt,boxsep=-1mm]
% \textbf{Answer to RQ3.}
% Hint utility is not intrinsic to the hint alone.
% Chat models can use compact hints in-context, while this specialized downstream decompilation backbone requires adaptation before the same semantic interface affects its outputs.
% \end{tcolorbox}

\section{Related Work}
\label{sec:related}

Existing work addresses adjacent parts of the \textsc{STILL} pipeline, 
including LLM-based decompilation and refinement~\citep{armengol2024slade,tan2024llm4decompile,exebench,shypula2026decaf,wang2025context}, binary type and data-structure recovery~\citep{lacomis2019dire,chen2022dirty,xie2024resym,tie2011,retypd2016,zhu2024tygr,typeforge2025,schwartz2018ooanalyzer}, and binary representation learning~\citep{ding2019asm2vec,wang2022jtrans,gu2022uniasm,peixoto2023ktrans}. We discuss these lines of work in Appendix~\ref{app:related}.
% where we also contrast them with our focus on recovering STL container semantics from stripped control-flow graphs and exposing the recovered semantics as compact hints for downstream LLM refinement.
% TIARA~\citep{wang2022tiara} is a related STL-specific type recovery work;
% Appendix~\ref{app:tiara} details its relationship to \textsc{STILL}.

\section{Discussion}
\label{sec:discussion}

\textsc{STILL} supports three lessons.
First, recoverability follows implementation regularity:
containers that leave stable layout and access residue in stripped binaries are 
much easier %FIXME careful.  consider "... that leave stable layout and access residue in stripped binaries can more accurately and faithfully recovered than containers that ..."
to recover than containers that collapse to shared low-level structures.
Second, STL decompilation failures are not monolithic:
the completed hint experiment %FIXME which one? refer to the subsection or RQ
isolates the part that can be 
repaired  % FIXME what "part" and why does it need "repairing"?
by restoring signatures and container-facing interfaces, while leaving other decompiler errors outside the claim.%FIXME I don't understand the rest of this sentence. "outside the claim"? which claim?  Why is this not monolithic?
Third, semantic hints are interfaces rather than facts alone.
DeepSeek-chat can use recovered hints directly in prompting, whereas LLM4Decompile-Ref needs format-matched adaptation before oracle or recovered hints translate into execution gains.

%These observations define the next experimental boundary for \textsc{STILL}.
%FIXME this sentence reads like some summary from an LLM.  Beware using the LLM to write whole sections for you.   As such, I advise removing this paragraph altogether. 
% The paper should not claim that all erased high-level semantics are recoverable from binaries.
%Instead, it shows a narrower and empirically useful claim:
%some STL semantics survive as compiler- and implementation-mediated residues, and exposing those residues through a structured hint channel can close a concrete decompilation failure path.

\section{Conclusion}
\label{sec:conclusion}

\textsc{STILL} shows that a major failure mode in LLM-assisted C++ decompilation is not simply weak raw decompilation, %FIXME I put "raw decompilation" because I'm not sure if "generation" is the right word to use -- decompilers don't really "generate" code... 
but the absence of source-level STL semantics after compilation, optimization, and stripping remove or obscure them.
Traditional decompiler output for stripped binaries often fails to recover these container abstractions, yet they remain partially recoverable as layout, access, and control-flow residue in the binary CFG.
By recovering and injecting this missing semantic channel, \textsc{STILL} improves executable decompilation and repairs many of the signature/container failures that motivate the method.
The results also show that semantic hints are interfaces: %FIXME I know this is used a few places throughout the paper, but I'm not sure it's consequential.  Why does it matter if these hints are "interfaces"?
some downstream models can consume them in context, while specialized decompilation models may require adaptation.
This points to a broader direction for LLM decompilation: LLM refiners
%should be 
can be paired with explicit semantic recovery for abstractions that compilation, optimization, and stripping remove or obscure before the LLM begins refining decompiled source. %FIXME I'm thinking "...erases before starting to refine decompiled source."

%FIXME broader notes here. 

% - I don't quite understand why it is important to note "interface" here and throughout the paper.  Part of the issue may stem from how this term is used elsewhere in literature.  This makes me think it's like a function that can be called (the "I" in API) that does something. 
% - I think we need to clarify the terms:
%    - generation     code generation is a step that the compiler does to generate assembly.  But I haven't really seen "generation" be used to describe the process of decompilation, unless perhaps when there's a language model involved.   That is what makes me think that your process of using an LLM to enhance the decompilation produced by Ghidra could be called "generation" (but more appropriately "refinement")
%    - decompilation  decompilation is the process of turning assembly code into source code.   
%    - refinement     I think I would call it "refinement" when you take raw decompilation from Ghidra and enhance it with semantic information. 

%    - One thing that's missing is "intermediate representation."   IR refer to things like control flow graphs, abstract syntax trees, etc. that are used in the compilation toolchain.  Compilers turn source code into IR and then assembly and binary.  Decompilers lift assembly into IR.  This may help you clarify parts of the methodology where you're describing looking through basic blocks in the cfg. 

% Limitations section is required by *ACL venues and does not count toward the
% 8-page main-text limit.
\section*{Limitations}
\label{sec:limitations}

\paragraph{Scope of semantic recovery.}
\textsc{STILL} is a controlled study of whether a small amount of missing STL
semantics can be recovered from stripped binaries and made useful to an LLM
refiner; it is not a complete C++ type-recovery system.  Its supervised
interface predicts function-level presence of only five containers
(\texttt{map}, \texttt{queue}, \texttt{set}, \texttt{string}, and
\texttt{vector}), rather than variable-level types, template arguments,
container nesting, iterator identities, or arbitrary library abstractions.
These omissions are consequential: \texttt{std::array} can lower to ordinary
stack or contiguous-memory accesses with little library-specific residue,
nested containers can leave evidence for multiple abstractions without
revealing their nesting relation, and optimized iterator code can resemble
pointer arithmetic and bounds checks.  The evidence is consequently strongest
for \texttt{string} and \texttt{vector}; estimates for the less frequent,
similarly implemented \texttt{map} and \texttt{set} classes require greater
caution.  Variable-level recovery could yield more targeted hints, but would
require variable-level annotations and reliable alignment among source
variables, binary addresses, and decompiled variables.

\paragraph{Toolchain and pipeline dependence.}
Our experiments use \texttt{g++} on x86-64 with libstdc++, angr
\texttt{CFGFast} for control-flow recovery, and Ghidra for the decompilation
context.  The residues learned in this setting may not transfer unchanged to
libc++, MSVC STL, other architectures, allocator configurations, or different
optimization pipelines.  Moreover, a correct function-level prediction does
not guarantee a useful refinement: missed CFG edges, type aliases, empty
Ghidra output, or a mismatch between binary structure and the decompiled
interface can still prevent recovery.  We do not evaluate direct
raw-assembly-to-source generation.

\paragraph{Strength of downstream evidence.}
$R_\text{exec}$ establishes that a reconstructed program compiles and passes
the benchmark test suite; it does not establish semantic equivalence to the
original C++ function.  A finite test suite samples only part of the input and
program-state space, and can therefore miss behavioral differences on
uncovered paths or boundary conditions.  Stronger evaluation should combine
differential testing of the reconstructed and reference programs on
independently generated inputs---including coverage-guided fuzzing to expand
input-space exploration---with bounded symbolic execution or equivalence
checking that searches for a counterexample under explicit assumptions about
the memory model and external-library behavior.  Fuzzing can increase
confidence but cannot prove equivalence, while symbolic methods are limited by
path explosion and incomplete modeling of C++ runtime and library code.

\paragraph{Potential risks and responsible use.}
Decompilation and binary analysis can be misused to inspect software without
authorization.  We intend \textsc{STILL} for authorized reverse engineering,
maintenance, security analysis, and research; users should respect software
licenses, terms of use, and applicable law.

% Acknowledgments may be added here for the camera-ready version.

% acl.sty already sets \bibliographystyle{acl_natbib}.
\bibliography{references}

@article{wang2025context,
  title={Context-Guided Decompilation: A Step Towards Re-executability},
  author={Wang, Xiaohan and Hu, Yuxin and Leach, Kevin},
  journal={arXiv preprint arXiv:2511.01763},
  year={2026},
  url={https://arxiv.org/abs/2511.01763}
}

@online{zhang2026ConstraintGuidedMultiAgent,
  title = {Constraint-Guided Multi-Agent Decompilation for Executable Binary Recovery},
  author = {Zhang, Yifan and Wang, Xiaohan and Zhang, Yueke and Huang, Yu and Leach, Kevin},
  date = {2026-05-01},
  eprint = {2604.23940},
  eprinttype = {arXiv},
  eprintclass = {cs.SE},
  doi = {10.48550/arXiv.2604.23940},
  url = {http://arxiv.org/abs/2604.23940},
  pubstate = {prepublished}
}

@misc{cao2023RevisitingDeepa,
  title = {Revisiting Deep Learning for Variable Type Recovery},
  author = {Cao, Kevin and Leach, Kevin},
  year = {2023},
  eprint = {2304.03854},
  archivePrefix = {arXiv},
  primaryClass = {cs.LG},
  doi = {10.48550/arXiv.2304.03854},
  url = {https://arxiv.org/abs/2304.03854}
}

@inproceedings{yang2025HumanStudya,
  title = {A Human Study of Automatically Generated Decompiler Annotations},
  booktitle = {2025 55th Annual IEEE/IFIP International Conference on Dependable Systems and Networks (DSN)},
  author = {Yang, Yuwei and Grandel, Skyler and Lacomis, Jeremy and Schwartz, Edward and Vasilescu, Bogdan and Le Goues, Claire and Leach, Kevin},
  year = {2025},
  pages = {129--142},
  issn = {2158-3927},
  doi = {10.1109/DSN64029.2025.00026},
  url = {https://ieeexplore.ieee.org/document/11068876},
  eventtitle = {2025 55th Annual IEEE/IFIP International Conference on Dependable Systems and Networks (DSN)}
}

@misc{she2024wadec,
  title={{WaDec}: Decompiling {WebAssembly} Using Large Language Model},
  author={She, Xinyu and Zhao, Yanjie and Wang, Haoyu},
  year={2024},
  eprint={2406.11346},
  archivePrefix={arXiv},
  primaryClass={cs.SE},
  url={https://arxiv.org/abs/2406.11346}
}

@misc{wang2026alt4decompile,
  title={{ALT4Decompile}: Inferring {C}-aligned Abstract Loop Tree for {LLM}-Based Binary Decompilation},
  author={Wang, Yongpan and Liu, Puzhuo and Xu, Xin and Li, Siyuan and Zheng, Yaowen and Gu, Xiaodong and Shen, Beijun},
  year={2026},
  eprint={2509.14646},
  archivePrefix={arXiv},
  primaryClass={cs.SE},
  url={https://arxiv.org/abs/2509.14646}
}

@misc{david2025smartcontract,
  title={Decompiling Smart Contracts with a Large Language Model},
  author={David, Isaac and Zhou, Liyi and Song, Dawn and Gervais, Arthur and Qin, Kaihua},
  year={2025},
  eprint={2506.19624},
  archivePrefix={arXiv},
  primaryClass={cs.CR},
  url={https://arxiv.org/abs/2506.19624}
}

@misc{zou2025dlift,
  title={{D-LiFT}: Improving {LLM}-based Decompiler Backend via Code Quality-driven Fine-tuning},
  author={Zou, Muqi and Cai, Hongyu and Wu, Hongwei and Basque, Zion Leonahenahe and Khan, Arslan and Celik, Z. Berkay and Tian, Dave and Bianchi, Antonio and Wang, Ruoyu and Xu, Dongyan},
  year={2025},
  eprint={2506.10125},
  archivePrefix={arXiv},
  primaryClass={cs.CR},
  doi={10.48550/arXiv.2506.10125},
  url={https://arxiv.org/abs/2506.10125}
}

@misc{feng2024sc2dec,
  title={Self-Constructed Context Decompilation with Fined-grained Alignment Enhancement},
  author={Feng, Yunlong and Teng, Dechuan and Xu, Yang and Mu, Honglin and Xu, Xiao and Qin, Libo and Zhu, Qingfu and Che, Wanxiang},
  year={2024},
  eprint={2406.17233},
  archivePrefix={arXiv},
  primaryClass={cs.SE},
  url={https://arxiv.org/abs/2406.17233}
}

@misc{achamyeleh2026helios,
  title={{HELIOS}: Hierarchical Graph Abstraction for Structure-Aware {LLM} Decompilation},
  author={Achamyeleh, Yonatan Gizachew and Thomare, Harsh and Al Faruque, Mohammad Abdullah},
  year={2026},
  eprint={2601.14598},
  archivePrefix={arXiv},
  primaryClass={cs.SE},
  url={https://arxiv.org/abs/2601.14598}
}

@misc{shang2025binmetric,
  title={{BinMetric}: A Comprehensive Binary Analysis Benchmark for Large Language Models},
  author={Shang, Xiuwei and Chen, Guoqiang and Cheng, Shaoyin and Wu, Benlong and Hu, Li and Li, Gangyang and Zhang, Weiming and Yu, Nenghai},
  year={2025},
  eprint={2505.07360},
  archivePrefix={arXiv},
  primaryClass={cs.SE},
  url={https://arxiv.org/abs/2505.07360}
}

@misc{jiang2025ir,
  title={Can Large Language Models Understand Intermediate Representations in Compilers?},
  author={Jiang, Hailong and Zhu, Jianfeng and Wan, Yao and Fang, Bo and Zhang, Hongyu and Jin, Ruoming and Guan, Qiang},
  year={2025},
  eprint={2502.06854},
  archivePrefix={arXiv},
  primaryClass={cs.LG},
  url={https://arxiv.org/abs/2502.06854}
}

@inproceedings{liu2025firmwarerenaming,
  title={Function Renaming in Reverse Engineering of Embedded Device Firmware with {ChatGPT}},
  author={Liu, Puzhuo and Di, Peng and Jiang, Yu},
  booktitle={Proceedings of the 1st {ACM SIGPLAN} International Workshop on Language Models and Programming Languages},
  pages={57--65},
  year={2025},
  publisher={ACM},
  doi={10.1145/3759425.3763387},
  url={https://doi.org/10.1145/3759425.3763387}
}

@misc{wong2023refining,
  title={Refining Decompiled {C} Code with Large Language Models},
  author={Wong, Wai Kin and Wang, Huaijin and Li, Zongjie and Liu, Zhibo and Wang, Shuai and Tang, Qiyi and Nie, Sen and Wu, Shi},
  year={2023},
  eprint={2310.06530},
  archivePrefix={arXiv},
  primaryClass={cs.SE},
  url={https://arxiv.org/abs/2310.06530}
}

@inproceedings{hu2024degpt,
  title={{DeGPT}: Optimizing Decompiler Output with {LLM}},
  author={Hu, Peiwei and Liang, Ruigang and Chen, Kai},
  booktitle={Proceedings 2024 Network and Distributed System Security Symposium},
  year={2024},
  publisher={Internet Society},
  doi={10.14722/ndss.2024.24401},
  url={https://doi.org/10.14722/ndss.2024.24401}
}

@article{wong2025decllm,
  title={{DecLLM}: {LLM}-Augmented Recompilable Decompilation for Enabling Programmatic Use of Decompiled Code},
  author={Wong, Wai Kin and Wu, Daoyuan and Wang, Huaijin and Li, Zongjie and Liu, Zhibo and Wang, Shuai and Tang, Qiyi and Nie, Sen and Wu, Shi},
  journal={Proceedings of the ACM on Software Engineering},
  volume={2},
  number={ISSTA},
  pages={1841--1864},
  year={2025},
  doi={10.1145/3728958},
  url={https://doi.org/10.1145/3728958}
}

@misc{feng2025refdecompile,
  title={{ReF Decompile}: Relabeling and Function Call Enhanced Decompile},
  author={Feng, Yunlong and Li, Bohan and Shi, Xiaoming and Zhu, Qingfu and Che, Wanxiang},
  year={2025},
  eprint={2502.12221},
  archivePrefix={arXiv},
  primaryClass={cs.SE},
  url={https://arxiv.org/abs/2502.12221}
}

@inproceedings{zhou2026fidelitygpt,
  title={{FidelityGPT}: Correcting Decompilation Distortions with Retrieval Augmented Generation},
  author={Zhou, Zhiping and Li, Xiaohong and Feng, Ruitao and Zhang, Yao and Li, Yuekang and Feng, Wenbu and Wang, Yunqian and Li, Yuqing},
  booktitle={Proceedings 2026 Network and Distributed System Security Symposium},
  year={2026},
  publisher={Internet Society},
  doi={10.14722/ndss.2026.230989},
  url={https://doi.org/10.14722/ndss.2026.230989}
}

@misc{cui2026pcodetrans,
  title={{PCodeTrans}: Translate Decompiled Pseudocode to Compilable and Executable Equivalent},
  author={Cui, Yuxin and Gao, Zeyu and He, Shuxian and Qin, Siliang and Zhang, Chao},
  year={2026},
  eprint={2603.14855},
  archivePrefix={arXiv},
  primaryClass={cs.SE},
  url={https://arxiv.org/abs/2603.14855}
}

@phdthesis{cifuentes1994reverse,
  title={Reverse Compilation Techniques},
  author={Cifuentes, Cristina},
  school={Queensland University of Technology},
  year={1994},
  url={https://decompilation.wiki/decompilers/dec-history-1/}
}

@inproceedings{shoshitaishvili2016angr,
  title={{SoK}: (State of) The Art of War: Offensive Techniques in Binary Analysis},
  author={Shoshitaishvili, Yan and Wang, Ruoyu and Salls, Christopher and Stephens, Nick and Polino, Mario and Dutcher, Andrew and Grosen, John and Feng, Siji and Hauser, Christophe and Kruegel, Christopher and Vigna, Giovanni},
  booktitle={IEEE Symposium on Security and Privacy (SP)},
  pages={138--157},
  year={2016},
  doi={10.1109/SP.2016.17},
  url={https://ieeexplore.ieee.org/document/7546492}
}

@inproceedings{schwartz2013native,
  title={Native x86 Decompilation Using Semantics-Preserving Structural Analysis and Iterative Control-Flow Structuring},
  author={Schwartz, Edward J. and Lee, JongHyup and Woo, Maverick and Brumley, David},
  booktitle={22nd USENIX Security Symposium (USENIX Security 13)},
  pages={353--368},
  year={2013},
  publisher={USENIX Association},
  url={https://www.usenix.org/conference/usenixsecurity13/technical-sessions/presentation/schwartz}
}

@inproceedings{yakdan2015nomoregotos,
  title={No More Gotos: Decompilation Using Pattern-Independent Control-Flow Structuring and Semantics-Preserving Transformations},
  author={Yakdan, Khaled and Eschweiler, Sebastian and Gerhards-Padilla, Elmar and Smith, Matthew},
  booktitle={Network and Distributed System Security Symposium (NDSS)},
  year={2015},
  doi={10.14722/ndss.2015.23185},
  url={https://www.ndss-symposium.org/wp-content/uploads/2017/09/11_4_2.pdf}
}

@inproceedings{basque2024sailr,
  title={Ahoy {SAILR}! There is No Need to {DREAM} of {C}: A Compiler-Aware Structuring Algorithm for Binary Decompilation},
  author={Basque, Zion Leonahenahe and Bajaj, Ati Priya and Gibbs, Wil and O'Kain, Jude and Miao, Derron and Bao, Tiffany and Doup{\'e}, Adam and Shoshitaishvili, Yan and Wang, Ruoyu},
  booktitle={33rd USENIX Security Symposium (USENIX Security 24)},
  pages={361--378},
  year={2024},
  publisher={USENIX Association},
  url={https://www.usenix.org/conference/usenixsecurity24/presentation/basque}
}

@inproceedings{lacomis2019dire,
  title={Dire: A neural approach to decompiled identifier naming},
  author={Lacomis, Jeremy and Yin, Pengcheng and Schwartz, Edward and Allamanis, Miltiadis and Le Goues, Claire and Neubig, Graham and Vasilescu, Bogdan},
  booktitle={2019 34th IEEE/ACM International Conference on Automated Software Engineering (ASE)},
  pages={628--639},
  year={2019},
  organization={IEEE}
}

@inproceedings{tan2024llm4decompile,
  title={{LLM4Decompile}: Decompiling Binary Code with Large Language Models},
  author={Tan, Hanzhuo and Luo, Qi and Li, Jing and Zhang, Yuqun},
  booktitle={Proceedings of the 2024 Conference on Empirical Methods in Natural Language Processing (EMNLP)},
  year={2024},
  url={https://aclanthology.org/2024.emnlp-main.203/}
}

@article{dramko2025idioms,
  title={Idioms: Neural Decompilation With Joint Code and Type Definition Prediction},
  author={Dramko, Luke and Le Goues, Claire and Schwartz, Edward J.},
  journal={arXiv preprint arXiv:2502.04536},
  year={2025},
  url={https://arxiv.org/abs/2502.04536}
}

@article{sk2decompile,
  title={SK2Decompile: LLM-based Two-Phase Binary Decompilation from Skeleton to Skin},
  author={Tan, Hanzhuo and Li, Weihao and Tian, Xiaolong and Wang, Siyi and Liu, Jiaming and Li, Jing and Zhang, Yuqun},
  journal={arXiv preprint arXiv:2509.22114},
  year={2025},
  url={https://arxiv.org/abs/2509.22114}
}

@misc{shypula2026decaf,
  title={{Decaf}: Improving Neural Decompilation with Automatic Feedback and Search},
  author={Shypula, Alexander and Bastani, Osbert and Schwartz, Edward},
  year={2026},
  eprint={2605.11501},
  archivePrefix={arXiv},
  primaryClass={cs.SE},
  doi={10.48550/arXiv.2605.11501},
  url={https://arxiv.org/abs/2605.11501}
}

@misc{armengol2024slade,
  title={{SLaDe}: A Portable Small Language Model Decompiler for Optimized Assembly},
  author={Armengol-Estap{\'e}, Jordi and Woodruff, Jackson and Cummins, Chris and O'Boyle, Michael F. P.},
  year={2024},
  eprint={2305.12520},
  archivePrefix={arXiv},
  primaryClass={cs.PL},
  doi={10.48550/arXiv.2305.12520},
  url={https://arxiv.org/abs/2305.12520}
}

@misc{hosseini2022btc,
  title={Beyond the {C}: Retargetable Decompilation using Neural Machine Translation},
  author={Hosseini, Iman and Dolan-Gavitt, Brendan},
  year={2022},
  eprint={2212.08950},
  archivePrefix={arXiv},
  primaryClass={cs.CR},
  doi={10.48550/arXiv.2212.08950},
  url={https://arxiv.org/abs/2212.08950}
}

@inproceedings{katz2018rnn,
  title={Using recurrent neural networks for decompilation},
  author={Katz, Deborah S and Ruchti, Jason and Schulte, Eric},
  booktitle={2018 IEEE 25th international conference on software analysis, evolution and reengineering (SANER)},
  pages={346--356},
  year={2018},
  organization={IEEE}
}

@misc{katz2019neural,
  title={Towards Neural Decompilation},
  author={Katz, Omer and Olshaker, Yuval and Goldberg, Yoav and Yahav, Eran},
  year={2019},
  eprint={1905.08325},
  archivePrefix={arXiv},
  primaryClass={cs.PL},
  doi={10.48550/arXiv.1905.08325},
  url={https://arxiv.org/abs/1905.08325}
}

@article{liang2021neutron,
  title={Neutron: An Attention-based Neural Decompiler},
  author={Liang, Ruigang and Cao, Ying and Hu, Peiwei and Chen, Kai},
  journal={Cybersecurity},
  volume={4},
  number={5},
  year={2021},
  doi={10.1186/s42400-021-00070-0},
  url={https://cybersecurity.springeropen.com/articles/10.1186/s42400-021-00070-0}
}

@inproceedings{zhang2021osprey,
  title={Osprey: Recovery of variable and data structure via probabilistic analysis for stripped binary},
  author={Zhang, Zhuo and Ye, Yapeng and You, Wei and Tao, Guanhong and Lee, Wen-chuan and Kwon, Yonghwi and Aafer, Yousra and Zhang, Xiangyu},
  booktitle={2021 IEEE Symposium on Security and Privacy (SP)},
  pages={813--832},
  year={2021},
  organization={IEEE}
}

@inproceedings{wang2022tiara,
  title={Recovering container class types in C++ binaries},
  author={Wang, Xudong and Xu, Xuezheng and Li, Qingan and Yuan, Mengting and Xue, Jingling},
  booktitle={2022 IEEE/ACM International Symposium on Code Generation and Optimization (CGO)},
  pages={131--143},
  year={2022},
  organization={IEEE},
  url={https://ieeexplore.ieee.org/document/9741274}
}

@inproceedings{sure2025benchmark,
  title={Benchmarking Binary Type Inference Techniques in Decompilers},
  author={Soni, Vedant and Dutcher, Audrey and Bao, Tiffany and Wang, Ruoyu},
  booktitle={Proceedings of the 2025 Workshop on Software Understanding and Reverse Engineering (SURE)},
  year={2025},
  doi={10.1145/3733822.3764675},
  url={https://sure-workshop.org/accepted-papers/2025/sure25-8.pdf}
}

@inproceedings{schwartz2018ooanalyzer,
  title={Using logic programming to recover c++ classes and methods from compiled executables},
  author={Schwartz, Edward J and Cohen, Cory F and Duggan, Michael and Gennari, Jeffrey and Havrilla, Jeffrey S and Hines, Charles},
  booktitle={Proceedings of the 2018 ACM SIGSAC Conference on Computer and Communications Security},
  pages={426--441},
  year={2018},
  url={https://dl.acm.org/doi/pdf/10.1145/3243734.3243793}
}

@inproceedings{mycroft1999type,
  title={Type-Based Decompilation (or Program Reconstruction via Type Reconstruction)},
  author={Mycroft, Alan},
  booktitle={Programming Languages and Systems, 8th European Symposium on Programming (ESOP)},
  pages={208--223},
  year={1999},
  publisher={Springer},
  url={https://www.cl.cam.ac.uk/~am21/research/decomp/}
}

@article{caballero2016type,
  title={Type Inference on Executables},
  author={Caballero, Juan and Lin, Zhiqiang},
  journal={ACM Computing Surveys},
  volume={48},
  number={4},
  pages={65:1--65:35},
  year={2016},
  doi={10.1145/2896499},
  url={https://dl.acm.org/doi/10.1145/2896499}
}

@inproceedings{tie2011,
  title={{TIE}: Principled Reverse Engineering of Types in Binary Programs},
  author={Lee, JongHyup and Avgerinos, Thanassis and Brumley, David},
  booktitle={Network and Distributed System Security Symposium (NDSS)},
  year={2011},
  url={https://www.ndss-symposium.org/ndss2011/tie-principled-reverse-engineering-of-types-in-binary-programs/}
}

@inproceedings{retypd2016,
  title={Polymorphic type inference for machine code},
  author={Noonan, Matt and Loginov, Alexey and Cok, David},
  booktitle={Proceedings of the 37th ACM SIGPLAN Conference on Programming Language Design and Implementation},
  pages={27--41},
  year={2016}
}

@inproceedings{chen2022dirty,
  title={Augmenting Decompiler Output with Learned Variable Names and Types},
  author={Chen, Qibin and Lacomis, Jeremy and Schwartz, Edward J. and Le Goues, Claire and Neubig, Graham and Vasilescu, Bogdan},
  booktitle={31st USENIX Security Symposium (USENIX Security 22)},
  pages={4327--4343},
  year={2022},
  publisher={USENIX Association},
  url={https://www.usenix.org/conference/usenixsecurity22/presentation/chen-qibin}
}

@inproceedings{xie2024resym,
  title={{ReSym}: Harnessing {LLMs} to Recover Variable and Data Structure Symbols from Stripped Binaries},
  author={Xie, Danning and Zhang, Zhuo and Jiang, Nan and Xu, Xiangzhe and Tan, Lin and Zhang, Xiangyu},
  booktitle={Proceedings of the 2024 ACM SIGSAC Conference on Computer and Communications Security (CCS)},
  pages={4554--4568},
  year={2024},
  doi={10.1145/3658644.3670340},
  url={https://www.cs.purdue.edu/homes/lintan/publications/resym-ccs24.pdf}
}

@misc{stride2024,
  title={{STRIDE}: Simple Type Recognition In Decompiled Executables},
  author={Green, Harrison and Schwartz, Edward J. and Le Goues, Claire and Vasilescu, Bogdan},
  year={2024},
  eprint={2407.02733},
  archivePrefix={arXiv},
  primaryClass={cs.CR},
  doi={10.48550/arXiv.2407.02733},
  url={https://arxiv.org/abs/2407.02733}
}

@inproceedings{stewart2025dragon,
  title={{DRAGON}: Predicting Decompiled Variable Data Types with Learned Confidence Estimates},
  author={Stewart, Caleb and Gaede, Rhonda K. and Kulick, Jeffrey H.},
  booktitle={Workshop on Binary Analysis Research (BAR)},
  year={2025},
  doi={10.14722/bar.2025.23025},
  url={https://www.ndss-symposium.org/wp-content/uploads/bar2025-final25.pdf}
}

@inproceedings{zhu2024tygr,
  title={{TYGR}: Type Inference on Stripped Binaries using Graph Neural Networks},
  author={Zhu, Chang and Li, Ziyang and Xue, Anton and Bajaj, Ati Priya and Gibbs, Wil and Liu, Yibo and Alur, Rajeev and Bao, Tiffany and Dai, Hanjun and Doup{\'e}, Adam and Naik, Mayur and Shoshitaishvili, Yan and Wang, Ruoyu and Machiry, Aravind},
  booktitle={33rd USENIX Security Symposium (USENIX Security 24)},
  pages={4283--4300},
  year={2024},
  publisher={USENIX Association},
  url={https://www.usenix.org/conference/usenixsecurity24/presentation/zhu-chang}
}

@inproceedings{bosamiya2025trex,
  title={{TRex}: Practical Type Reconstruction for Binary Code},
  author={Bosamiya, Jay and Woo, Maverick and Parno, Bryan},
  booktitle={34th USENIX Security Symposium (USENIX Security 25)},
  year={2025},
  publisher={USENIX Association},
  url={https://www.microsoft.com/en-us/research/publication/trex-practical-type-reconstruction-for-binary-code/}
}

@inproceedings{typeforge2025,
  title={{TypeForge}: Synthesizing and Selecting Best-Fit Composite Data Types for Stripped Binaries},
  author={Wang, Yanzhong and Liang, Ruigang and Li, Yilin and Hu, Peiwei and Chen, Kai and Zhang, Bolun},
  booktitle={2025 IEEE Symposium on Security and Privacy (SP)},
  pages={1--18},
  year={2025},
  doi={10.1109/SP61157.2025.00193},
  url={https://jglobal.jst.go.jp/en/public/202502251947432038}
}

@inproceedings{schlichtkrull2018rgcn,
  title={Modeling relational data with graph convolutional networks},
  author={Schlichtkrull, Michael and Kipf, Thomas N and Bloem, Peter and Van Den Berg, Rianne and Titov, Ivan and Welling, Max},
  booktitle={European semantic web conference},
  pages={593--607},
  year={2018},
  organization={Springer}
}

@inproceedings{ding2019asm2vec,
  title={{Asm2Vec}: Boosting Static Representation Robustness for Binary Clone Search against Code Obfuscation and Compiler Optimization},
  author={Ding, Steven H. H. and Fung, Benjamin C. M. and Charland, Philippe},
  booktitle={2019 IEEE Symposium on Security and Privacy (SP)},
  pages={472--489},
  year={2019},
  organization={IEEE},
  url={https://ieeexplore.ieee.org/document/8835340}
}

@article{gu2022uniasm,
  title={Uniasm: Binary code similarity detection without fine-tuning},
  author={Gu, Yeming and Shu, Hui and Kang, Fei and Hu, Fan},
  journal={Neurocomputing},
  volume={630},
  pages={129646},
  year={2025},
  publisher={Elsevier},
  url={https://www.sciencedirect.com/science/article/pii/S0925231225003182?casa_token=i3Mj8hwDNQcAAAAA:F63kusHxBv2NhdNmeJurd7_BA2JQ6cspewgFQMcPNdONV8uoSJRyO_J9DpRKFGWfDEi_TFs}
}

@inproceedings{wang2022jtrans,
  title={{jTrans}: Jump-aware Transformer for Binary Code Similarity Detection},
  author={Wang, Hao and Qu, Wenjie and Katz, Gilad and Zhu, Wenyu and Gao, Zeyu and Qiu, Han and Zhuge, Jianwei and Zhang, Chao},
  booktitle={Proc.\ ISSTA},
  year={2022},
  url={https://conf.researchr.org/details/issta-2022/issta-2022-technical-papers/4/jTrans-Jump-Aware-Transformer-for-Binary-Code-Similarity-Detection}
}

@misc{peixoto2023ktrans,
  title={{kTrans}: Knowledge-Aware Transformer for Binary Code Embedding},
  author={Zhu, Wenyu and Wang, Hao and Zhou, Yuchen and Wang, Jiaming and Sha, Zihan and Gao, Zeyu and Zhang, Chao},
  year={2023},
  eprint={2308.12659},
  archivePrefix={arXiv},
  primaryClass={cs.SE},
  doi={10.48550/arXiv.2308.12659},
  url={https://arxiv.org/abs/2308.12659}
}

@article{li2022alphacode,
  title={Competition-level code generation with alphacode},
  author={Li, Yujia and Choi, David and Chung, Junyoung and Kushman, Nate and Schrittwieser, Julian and Leblond, R{\'e}mi and Eccles, Tom and Keeling, James and Gimeno, Felix and Dal Lago, Agustin and others},
  journal={Science},
  volume={378},
  number={6624},
  pages={1092--1097},
  year={2022},
  publisher={American Association for the Advancement of Science}
}

@misc{hu2021lora,
  author = {Hu, Edward J. and Shen, Yelong and Wallis, Phillip and Allen-Zhu, Zeyuan and Li, Yuanzhi and Wang, Shean and Wang, Lu and Chen, Weizhu},
  title = {{LoRA}: Low-Rank Adaptation of Large Language Models},
  publisher = {arXiv},
  year = {2021},
  doi = {10.48550/arXiv.2106.09685},
  url = {https://arxiv.org/abs/2106.09685}
}

@inproceedings{wu2015libraries,
  title={How Do Developers Use {C++} Libraries? An Empirical Study},
  author={Wu, Di and Chen, Lin and Zhou, Yuming and Xu, Baowen},
  booktitle={Proceedings of the 27th International Conference on Software Engineering and Knowledge Engineering (SEKE)},
  year={2015},
  doi={10.18293/SEKE2015-009},
  url={https://ksiresearch.org/seke/seke15paper/seke15paper_9.pdf}
}

@inproceedings{exebench,
  title={{ExeBench}: An {ML}-Scale Dataset of Executable {C} Functions},
  author={Armengol-Estap{\'e}, Jordi and Woodruff, Jackson and Brauckmann, Alexander and de Souza Magalh{\~a}es, Jos{\'e} Wesley and O'Boyle, Michael F. P.},
  booktitle={Proceedings of the 6th ACM SIGPLAN International Symposium on Machine Programming (MAPS)},
  pages={50--59},
  year={2022},
  doi={10.1145/3520312.3534867},
  url={https://github.com/jordiae/exebench}
}

%%%%%%%%%%%%%%%%%%%%%%%%%%%%%%%%%%%%%%%%%%%%%%%%%%%%%%%%%%%%

\appendix
\section*{Appendix}

\section{Related Work}
\label{app:related}

\paragraph{LLM decompilation and refinement.}
Earlier neural decompilers cast source recovery as sequence-to-sequence translation
\citep{katz2018rnn,katz2019neural,liang2021neutron,hosseini2022btc}.
Recent systems extend this formulation with code LLMs, executable-function corpora, contextual information, and feedback-guided search
\citep{armengol2024slade,exebench,tan2024llm4decompile,dramko2025idioms,sk2decompile,shypula2026decaf}.
Recent work follows two directions. Direct systems translate a binary or intermediate representation into source, including WebAssembly and EVM bytecode, and improve generation with constructed recompilation contexts, statement alignment, or explicit tree structure
\citep{she2024wadec,david2025smartcontract,feng2024sc2dec,wang2026alt4decompile,feng2025refdecompile}.
Post-hoc systems instead refine decompiler pseudocode through readability optimization, recompilation and execution feedback, quality-guided fine-tuning, contextual or retrieved information, and repair loops; DIRTY is an earlier learned postprocessor for variable names and types
\citep{wong2023refining,hu2024degpt,wong2025decllm,zou2025dlift,wang2025context,zhang2026ConstraintGuidedMultiAgent,zhou2026fidelitygpt,cui2026pcodetrans,chen2022dirty}.
Both directions require the LLM to infer missing library abstractions from the binary-derived representation or repair feedback. \textsc{STILL} instead predicts an STL-specific prior from the stripped CFG before refinement, providing the LLM with a compact semantic constraint.

\paragraph{Structure, representations, and evaluation.}
Several studies likewise show that the representation supplied to an LLM is consequential:
HELIOS serializes hierarchical control-flow and call-graph structure for prompting,
and an empirical study examines how LLMs understand compiler intermediate representations
\citep{achamyeleh2026helios,jiang2025ir}.
BinMetric evaluates LLMs across binary-analysis tasks, while function renaming in firmware demonstrates a complementary use of LLMs for recovering human-facing binary semantics
\citep{shang2025binmetric,liu2025firmwarerenaming}.
Human studies further assess whether automatically generated annotations improve comprehension for reverse engineers
\citep{yang2025HumanStudya}.
Our use of an edge-typed CFG is different: it supports a supervised predictor of STL container labels, whose short output is then passed to the downstream refiner.

\paragraph{Binary type and data-structure recovery.}
Recovering high-level structure from binaries is a long-standing reverse-engineering problem,
from type-based decompilation and binary type inference
\citep{mycroft1999type,caballero2016type,tie2011,retypd2016}
to modern systems for variables, data structures, names, and source-level types in stripped or decompiled binaries
\citep{lacomis2019dire,zhang2021osprey,chen2022dirty,xie2024resym,stride2024,zhu2024tygr,bosamiya2025trex,typeforge2025,sure2025benchmark}.
Recent work also revisits learned variable type recovery from decompiler outputs
\citep{cao2023RevisitingDeepa}.
C++ template libraries make this problem sharper because object-oriented recovery tools such as OOAnalyzer~\citep{schwartz2018ooanalyzer}
do not identify which recovered abstractions are STL containers,
and STL/template reverse engineering has largely remained a specialized analysis problem~\citep{wu2015libraries}.
\paragraph{TIARA and \textsc{STILL}.}
\label{app:tiara}

TIARA is the closest STL-specific recovery system, but it operates under a different input assumption and prediction granularity from \textsc{STILL} (Table~\ref{tab:tiara-still}).  TIARA receives a known variable address and returns a variable-level container type, whereas \textsc{STILL} consumes a whole stripped-function CFG and returns the function-level set of container labels to be rendered as an LLM hint.  Consequently, the two systems cannot be compared through a single direct number: their query units, targets, and evaluation endpoints differ.  DRAGON, TYGR, and TRex likewise address variable/decompiled-type inference or reconstruction rather than an LLM-facing function-level semantic interface~\citep{stewart2025dragon,zhu2024tygr,bosamiya2025trex}.  Appendix~\ref{app:related} provides the broader comparison.

\begin{table}[!t]
    \centering
    \small
    \caption{TIARA--\textsc{STILL} comparison.  The systems are complementary rather than directly numerically comparable.}
    \label{tab:tiara-still}
    \setlength{\tabcolsep}{3pt}
    \renewcommand{\arraystretch}{1.05}
    \resizebox{\linewidth}{!}{%
    \begin{tabular}{lp{0.31\linewidth}p{0.31\linewidth}}
        \toprule
        Dimension & TIARA & \textsc{STILL} \\
        \midrule
        Input & Known variable address in a C++ binary & Whole stripped-function CFG \\
        Output & Variable-level container type & Function-level container-label set \\
        Goal & Static binary type inference & Semantic hints for LLM-assisted decompilation \\
        Evaluation & Variable-level precision, recall, and F1 & Container recovery and downstream $R_\text{exec}$ \\
        \bottomrule
    \end{tabular}%
    }
\end{table}

\paragraph{Binary representation learning.}
Traditional decompilers use control-flow recovery, structuring, and IR-level analysis to produce readable decompiled source
\citep{cifuentes1994reverse,schwartz2013native,yakdan2015nomoregotos,shoshitaishvili2016angr,basque2024sailr},
but these analyses do not restore STL container semantics (let alone types, classes, or structures from other libraries) once template instantiations have been inlined and symbols stripped.
Binary representation learning provides reusable encoders for instruction and CFG regularities, including assembly embeddings and control-flow-sensitive models for similarity and clone-search tasks
\citep{ding2019asm2vec,wang2022jtrans,gu2022uniasm,peixoto2023ktrans}.
We use edge-typed CFG message passing~\citep{schlichtkrull2018rgcn} as an implementation dependency, not as the main contribution.
We instead predict container-level STL use from low-level CFG and memory-access evidence, then expose those predictions as compact hints for downstream LLM refinement.

\section{Dataset and Audit Details}
\subsection{STL Container Usage Survey}
\label{app:stl-survey}

To quantify the prevalence of STL containers in real-world C++ projects, we surveyed the 100 most-starred C++ repositories on GitHub (all with $>$1,000 stars).
For each repository, we sampled 10 source files (prioritizing \texttt{src/} and \texttt{include/} directories) and searched for \texttt{\#include} directives of 11 STL container headers: \texttt{vector}, \texttt{string}, \texttt{map}, \texttt{set}, \texttt{unordered\_map}, \texttt{unordered\_set}, \texttt{list}, \texttt{deque}, \texttt{array}, \texttt{queue}, and \texttt{stack}.

\paragraph{Results.}
Under this conservative 10-file sampling, 76 of 100 repositories (76.0\%) contained at least one STL container header.
Manual verification of the 24 undetected repositories revealed that 17 are confirmed STL users whose headers were missed due to deep directory structures (e.g., \texttt{microsoft/terminal}), project-specific header wrappers (e.g., \texttt{duckdb} wraps \texttt{std::vector} in \texttt{duckdb/common/vector.hpp}), or Qt-based codebases that co-use STL alongside Qt containers.
The remaining 7 repositories are tutorials, build tooling, embedded/C-style projects, or legacy code.
This yields a corrected estimate of \textbf{93/100 (93.0\%)} repositories using STL containers.

\paragraph{Per-header breakdown.}
Table~\ref{tab:stl-survey} shows the detection rate for each header under the 10-file sampling (lower bound).
\texttt{string} (57.0\%) and \texttt{vector} (56.0\%) are the most prevalent.

\begin{table}[!t]
    \caption{STL container header prevalence among the top 100 most-starred C++ repositories on GitHub (10-file sampling, lower bound).}
    \label{tab:stl-survey}
    \centering
    \resizebox{\linewidth}{!}{%
    \begin{tabular}{lcc}
        \toprule
        Header & Repos & \% \\
        \midrule
        \texttt{string} & 57 & 57.0 \\
        \texttt{vector} & 56 & 56.0 \\
        \texttt{map} & 23 & 23.0 \\
        \texttt{unordered\_map} & 19 & 19.0 \\
        \texttt{set} & 16 & 16.0 \\
        \texttt{list} & 11 & 11.0 \\
        \texttt{unordered\_set} & 10 & 10.0 \\
        \texttt{array} & 10 & 10.0 \\
        \texttt{deque} & 5 & 5.0 \\
        \texttt{queue} & 4 & 4.0 \\
        \texttt{stack} & 1 & 1.0 \\
        \midrule
        \textit{Any container} & \textit{76 (93$^*$)} & \textit{76.0 (93.0$^*$)} \\
        \bottomrule
        \multicolumn{3}{l}{\footnotesize $^*$After manual verification of false negatives.}
    \end{tabular}%
    }
\end{table}

\subsection{Residual Symbol Audit}
\label{app:symbol-audit}

Although the binaries are stripped, a small number of external library-call names can remain in the assembly.
We audit whether these names directly reveal the true STL container type.
Table~\ref{tab:symbol-audit} shows that such direct name cues are rare overall and concentrated in \texttt{string}.

\begin{table}[!t]
    \centering
    \small
    \caption{Audit of residual external-call names in stripped assembly. We count records where a remaining external-call name directly indicates the true STL container type.}
    \label{tab:symbol-audit}
    \resizebox{\linewidth}{!}{%
    \begin{tabular}{lcc}
        \toprule
        Class & Records with direct name cue / total & Percentage \\
        \midrule
        \texttt{map} & 8 / 2{,}043 & 0.4 \\
        \texttt{queue} & 0 / 1{,}499 & 0.0 \\
        \texttt{set} & 4 / 1{,}845 & 0.2 \\
        \texttt{string} & 215 / 3{,}201 & 6.7 \\
        \texttt{vector} & 21 / 5{,}053 & 0.4 \\
        \midrule
        Overall & 218 / 8{,}900 & 2.4 \\
        \bottomrule
    \end{tabular}%
    }
\end{table}

\section{Experimental Details}
\subsection{Controlled Toolchain and Pipeline}
\label{app:toolchain}

Table~\ref{tab:toolchain} specifies the pipeline used for the reported results.
The results are evidence within this controlled g++/x86-64/libstdc++ setting,
not a claim of unchanged performance across decompilers, standard-library
implementations, architectures, or compilation pipelines.

\begin{table}[!t]
    \centering
    \small
    \caption{Toolchain and pipeline for the controlled experiments.}
    \label{tab:toolchain}
    \setlength{\tabcolsep}{3pt}
    \renewcommand{\arraystretch}{1.05}
    \resizebox{\linewidth}{!}{%
    \begin{tabular}{lp{0.64\linewidth}}
        \toprule
        Component & Configuration \\
        \midrule
        Source and platform & CodeContests and HumanEval-C++ functions; x86-64; libstdc++ \\
        Compilation & \texttt{g++} 11.4.0 with \texttt{-g}; O0, O1, O2, and O3; debug metadata is used only for source-to-binary alignment \\
        Stripping & Symbols are stripped from worker copies before analysis; model inputs never include debug metadata \\
        CFG extraction & angr 9.2.193, \texttt{CFGFast}, with typed control-flow edges and stripped-assembly residue features \\
        Decompilation & Ghidra decompiler; the version is pinned and recorded in the artifact environment manifest \\
        Downstream protocol & Raw Ghidra output plus the documented no-hint or container-hint prompt; compile and unit-test evaluation \\
        \bottomrule
    \end{tabular}%
    }
\end{table}

\subsection{Contamination Control: Fresh Synthetic Evaluation}
\label{app:contamination-control}

To test whether the results depend on potentially pretraining-contaminated benchmarks, we construct a fresh synthetic C++17 evaluation set locally with fixed seed \texttt{20260710}.  It contains 125 functions: 100 STL-bearing functions and 25 no-STL hard negatives.  Twenty of 25 auditable template families exercise \texttt{map}, \texttt{queue}, \texttt{set}, \texttt{string}, and \texttt{vector}, alone or in multi-label and nested combinations; the other five cover scalar integer arithmetic, raw C-style arrays, bitwise operations, null-terminated C-style character arrays, and fixed-size two-dimensional arrays.  Each family yields five variants through constants, window sizes, thresholds, and test cases, and its labels are derived directly from the template and verified against the generated function body.

All 125 source functions compile and pass their source-level tests before inclusion.  We compile each at O0--O3 and retain all 500 resulting records after symbol stripping, CFG extraction, and Ghidra decompilation.  Normalized-source deduplication finds no exact duplicate within this set or against our local HumanEval and CodeContests copies.  The set was not used for training, prompt selection, threshold tuning, or model selection.  Across source functions, the label counts are \texttt{map}=35, \texttt{queue}=30, \texttt{set}=35, \texttt{string}=40, and \texttt{vector}=60.

\begin{table}[!t]
    \centering
    \small
    \caption{Fresh synthetic-set performance of the frozen semantic extractor (500 retained binary records).}
    \label{tab:synthetic-extractor}
    \setlength{\tabcolsep}{4pt}
    \renewcommand{\arraystretch}{1.05}
    \resizebox{\linewidth}{!}{%
    \begin{tabular}{lrrrr}
        \toprule
        Split & $n$ & Label accuracy & Macro-F1 & Gate F1 \\
        \midrule
        Overall & 500 & 87.8 & 81.0 & 97.2 \\
        O0 & 125 & 79.0 & 68.9 & 92.3 \\
        O1 & 125 & 91.5 & 85.8 & 97.4 \\
        O2 & 125 & 90.6 & 83.7 & 99.5 \\
        O3 & 125 & 89.9 & 84.4 & 99.5 \\
        \bottomrule
    \end{tabular}%
    }
\end{table}

The frozen extractor achieves 81.0 macro-F1 and 97.2 gate F1 overall (Table~\ref{tab:synthetic-extractor}).  For a representative retained O2 record whose source uses \texttt{vector<string>}, \texttt{map<string,int>}, and string-prefix operations, it predicts exactly \texttt{map}, \texttt{string}, and \texttt{vector}.  Without a hint, the downstream refiner instead infers a \texttt{std::pair<std::string*, std::string*>*} interface and fails to compile; with this type-only hint it recovers a \texttt{std::vector<std::string>\&} parameter and a \texttt{std::map<std::string,int>} prefix-counting implementation that passes both generated tests.

\begin{table}[!t]
    \centering
    \small
    \caption{Fresh synthetic downstream evaluation with DeepSeek-chat.  Entries are percentages except edit similarity.}
    \label{tab:synthetic-downstream}
    \setlength{\tabcolsep}{4pt}
    \renewcommand{\arraystretch}{1.05}
    \resizebox{\linewidth}{!}{%
    \begin{tabular}{lrrrr}
        \toprule
        Condition & $n$ & $R_\text{comp}$ & $R_\text{exec}$ & Edit similarity \\
        \midrule
        No hint & 500 & 27.4 & 22.8 & 0.120 \\
        Type-only hint & 500 & 52.6 & 36.2 & 0.144 \\
        No hint, STL-bearing & 400 & 13.5 & 8.5 & 0.075 \\
        Type-only hint, STL-bearing & 400 & 46.2 & 26.2 & 0.114 \\
        \bottomrule
    \end{tabular}%
    }
\end{table}

On the same held-out records, type-only hints improve overall $R_\text{exec}$ from 22.8\% to 36.2\% and STL-bearing $R_\text{exec}$ from 8.5\% to 26.2\% (Table~\ref{tab:synthetic-downstream}).  These controls provide stronger evidence on freshly generated data, although they cannot rule out all broader forms of pretraining contamination.

\section{Semantic Extraction Details}
\subsection{RQ1 Per-Class F1}
\label{app:rq1-perclass}

Table~\ref{tab:main-perclass} extends the headline RQ1 results (Table~\ref{tab:rq1-semantic-extraction}) with per-class F1 on the 5 container classes.

\begin{table}[!t]
    \caption{Per-class F1 of the selected RQ1 semantic extractor (3-layer RGCN + residual + LayerNorm, per-class attention pooling, continuous+tree residue features, seed 44) on the in-domain CodeContests stripped held-out set, complementing the aggregate Macro-F1/Gate columns reported in Table~\ref{tab:rq1-semantic-extraction}.}
    \label{tab:main-perclass}
    \centering\small
    \resizebox{\linewidth}{!}{%
    \begin{tabular}{lccccc}
        \toprule
        Opt & map & queue & set & string & vector \\
        \midrule
        O0 & 0.608 & 0.720 & 0.530 & 0.779 & 0.771 \\
        O1 & 0.840 & 0.907 & 0.669 & 0.916 & \textbf{0.843} \\
        O2 & 0.832 & \textbf{0.944} & \textbf{0.749} & 0.899 & 0.831 \\
        O3 & \textbf{0.850} & 0.926 & 0.729 & \textbf{0.924} & 0.820 \\
        \bottomrule
    \end{tabular}%
    }
\end{table}

\subsection{Semantic Extractor Implementation Details}
\label{app:extractor-implementation}

The selected semantic extractor computes all residue features per basic block
from stripped assembly parsed from the recovered CFG.
It does not use source-level labels, debug metadata, Ghidra decompiler text,
or residual symbol names as node features.

\textsc{ContRes} is a 33-dimensional block-local vector
for contiguous-container evidence.
It includes log-scaled counts of non-stack memory references, loads, stores,
distinct field offsets, offset span, and offset bins at 0, 8, 16, 24, and 32+.
It also includes binary indicators for 0/8/16 and 0/8/16/24 layout tuples,
memory-width counts and byte-access ratios,
stride and indexing cues from addressing scales, add-by-1/4/8, shifts, and pointer differences,
and capacity/length, null-store, copy-call, and copy-loop predicates.

\textsc{TreeRes} is a 68-dimensional vector
for ordered-container evidence.
Its block-local features measure pointer loads and stores, base-register and offset diversity,
parent/left/right-like offset patterns, color-byte evidence,
memory comparisons followed by branches, pointer-chasing scores, and branch density.
For map/set disambiguation, it records payload offsets at 32+, payload loads and stores,
qword payload access, compare-to-payload proximity, pair-second access,
key-value offset predicates, and a map-vs-set margin.
Because tree traversals often span multiple basic blocks,
\textsc{TreeRes} also appends CFG-context summaries computed over each node's two-hop forward region,
its strongly connected component, and the whole graph.
These region summaries are written back to the node
before RGCN message passing.

\section{Artifact and Reproducibility}
\label{app:artifact}

The artifact contains the versioned data used in the reported experiments and an end-to-end reproduction guide.  Its README specifies the CodeContests source selection, preprocessing and label-generation steps, deterministic split generation, the compiler and dependency environment, and the commands that construct data, train the extractor, and run the recovery and downstream evaluations.  It also documents the prompt and hint formats, evaluation scripts, expected metric files and tables, and the planned public \textsc{StlBench} release.

For the headline semantic extractor, the artifact records the continuous+tree residue representation, a 3-layer RGCN with 128 hidden dimensions, residual connections, LayerNorm, per-class attention pooling, 30 epochs of Adam at learning rate $10^{-3}$, and seed~44.  It records all random seeds and the exact package/compiler versions used for each run, including g++~11.4.0 and angr~9.2.193 in the controlled setting of this paper; the Ghidra version is pinned in the environment manifest.  The downstream documentation gives the LoRA settings, test harness, prompts, and expected compile/execution outputs used for the reported conditions.

\end{document}